\documentclass[11pt]{article}
\usepackage{graphicx,amsmath,amsfonts,bm,cite,epsfig,epsf,url,color,subfigure}
\usepackage{fullpage}
\usepackage[small,bf]{caption}
\usepackage{times}
\usepackage{algorithmic,algorithm}
\newtheorem{theorem}{Theorem}[section]

\newtheorem{corollary}[theorem]{Corollary}

\newcommand{\C}{\mathbb{C}}

\newcommand{\Z}{\mathbb{Z}}
\newcommand{\<}{\langle}
\renewcommand{\>}{\rangle}

\newcommand{\cA}{\mathcal{A}}

\newcommand{\mapA}{\mathbb{A}}

\newcommand{\rank}{\operatorname{rank}}

\newcommand{\trace}{\operatorname{Tr}}

\numberwithin{equation}{section}

\def \endprf{\hfill {\vrule height6pt width6pt depth0pt}\medskip}
\newenvironment{proof}{\noindent {\bf Proof} }{\endprf\par}

\newcommand{\dirfigs}{figures}
\renewcommand{\epsilon}{\varepsilon}

\title{Phase Retrieval via Matrix Completion}

\author{Emmanuel J. Cand\`{e}s\thanks{Departments of Mathematics and
    of Statistics, Stanford University, Stanford CA 94305}, \, Yonina
  C. Eldar\thanks{Department of Electrical Engineering Technion, Israel
    Institute of Technology, Israel}, \, Thomas
  Strohmer\thanks{Department of Mathematics, University of California
    at Davis, Davis CA}\,\, and Vladislav Voroninski\thanks{Department of
    Mathematics, University of California at Berkeley, Berkeley CA}}

\date{August 2011}

\begin{document}

\maketitle

\vspace{-0.3in}

\begin{abstract}
  This paper develops a novel framework for phase retrieval, a problem
  which arises in X-ray crystallography, diffraction imaging,
  astronomical imaging and many other applications. Our approach, 
  called PhaseLift,
  combines multiple structured illuminations together with ideas from
  convex programming to recover the phase from intensity measurements,
  typically from the modulus of the diffracted wave.  We demonstrate
  empirically that any complex-valued object can be recovered from the
  knowledge of the magnitude of just a few diffracted patterns by
  solving a simple convex optimization problem inspired by the recent
  literature on matrix completion. More importantly, we also
  demonstrate that our noise-aware algorithms are stable in the sense
  that the reconstruction degrades gracefully as the signal-to-noise
  ratio decreases. Finally, we introduce some theory showing that one
  can design very simple structured illumination patterns such that
  three diffracted figures uniquely determine the phase of the object
  we wish to recover.
  \end{abstract}


\section{Introduction}
\label{sec:intro}

\subsection{The phase retrieval problem}

This paper considers the fundamental problem of recovering a general
signal, an image for example, from the magnitude of its Fourier
transform. This problem, also known as {\em phase retrieval}, arises
in many applications and has challenged engineers, physicists, and
mathematicians for decades.  Its origin comes from the fact
that detectors can often times only record the squared modulus of the
Fresnel or Fraunhofer diffraction pattern of the radiation that is
scattered from an object. In such settings, one cannot measure the
phase of the optical wave reaching the detector and, therefore, much
information about the scattered object or the optical field is lost
since, as is well known, the phase encodes a lot of the structural
content of the image we wish to form. 

Historically, the first application of phase retrieval is X-ray
crystallography \cite{Mil90,Har93}, and today this may still very well
be the most important application.  Over the last century or so, this
field has developed a wide array of techniques to recover Bragg peaks
from missing-phase data.  Of course, the phase retrieval problem
permeates many other areas of imaging science, and other applications
include diffraction imaging \cite{Bun07}, optics \cite{Wal63},
astronomical imaging \cite{DF87}, microscopy~\cite{MIS08}, to name
just a few.  In particular, X-ray tomography has become an invaluable
tool in biomedical imaging to generate quantitative 3D density maps of
extended specimens on the nanoscale~\cite{Die10}.  Other subjects
where phase retrieval plays an important role are quantum mechanics
\cite{Rei44,Cor06} and even differential geometry \cite{BSV02}.  We
note that phase retrieval has seen a resurgence in activity in recent
years, fueled on the one hand by the desire to image individual
molecules and other nano-particles, and on the other hand by new
imaging capabilities: one such recent modality is the availability of
new X-ray synchrotron sources that provide extraordinary X-ray fluxes,
see for example \cite{Mil06,Sca06,Bog08,MIS08,Die10}.  References and
various instances of the phase retrieval problem as well as some
theoretical and numerical solutions can be found
in~\cite{Hur89,LBL02,KST95}.

There are many ways in which one can pose the phase-retrieval problem,
for instance depending upon whether one assumes a continuous or
discrete-space model for the signal. In this paper, we consider finite
length signals (one-dimensional or multi-dimensional) for simplicity, 
and because numerical algorithms ultimately operate with digital data.
To fix ideas, suppose we have a 1D signal $x = (x[0], x[1],$ $\ldots,
x[n-1]) \in \C^n$ and write its Fourier transform as
\begin{equation}
  \label{eq:fourier}
  {\hat x}[\omega] = 
  \frac{1}{\sqrt{n}} \sum_{0 \le t < n} x[t] e^{-i 2\pi \omega t/n}, 
\quad \omega \in \Omega. 
\end{equation}
Here, $\Omega$ is a grid of sampled frequencies and an important
special case is $\Omega = \{0,1, \ldots, n-1\}$ so that the mapping is
the classical unitary discrete Fourier transform (DFT)\footnote{For
  later reference, we denote the Fourier transform operator by $F$ and
  the inverse Fourier transform by $F^{-1}$.}.  The phase retrieval
problem consists in finding $x$ from the magnitude coefficients
$|{\hat x}[\omega]|$, $\omega \in \Omega$.  When $\Omega$ is the usual
frequency grid as above and without further information about the
unknown signal $x$, this problem is ill-posed since there are many
different signals whose Fourier transforms have the same
magnitude. Clearly, if $x$ is a solution to the phase retrieval
problem, then (1) $c x$ for any scalar $c \in \C$ obeying $|c| = 1$ is
also solution, (2) the ``mirror function'' or time-reversed signal
$\bar{x}[-t \text{ mod } n]$ where $t = 0,1, \ldots, n-1$ is also
solution, and (3) the shifted signal $x[t - a \text{ mod } n]$ is also
a solution. From a physical viewpoint these ``trivial associates'' of
$x$ are acceptable ambiguities. But in general infinitely many
solutions can be obtained from $\{|\hat{x}[\omega]| : \omega \in
\Omega\}$ beyond these trivial associates \cite{San85}.

% The situation is more favorable for
% multidimensional---e.g.~two-dimensional---signals of the form $x =
% \{x[t_1, t_2]\}$ with $0 \le t_1 < n_1$ and $0 \le t_2 < n_2$, whose
% Fourier transform is given by
% \[ {\hat x}[\omega_1,\omega_2] = \frac{1}{\sqrt{n_1 n_2}} \sum x[t_1,
% t_2] e^{-i 2\pi (\omega_1 t_1/n_1 + \omega_2 t_2/n_2)}, \quad \omega =
% (\omega_1, \omega_2) \in \Omega.
% \]
% In this setting, it has been shown that twofold
% oversampling\footnote{This means that the grid is of the form
%   $\Omega_1 \times \Omega_2$ where $\Omega_i = \{0,1/2,1,3/2,\ldots,
%   n_i + 1/2\}$.}  in each dimension yields uniqueness for most
% finitely supported signals \cite{Hay82,San85}. Theoretical uniqueness
% conditions do not readily translate into numerical computability of
% the solution or in stable recovery from noisy data, a fact we shall
% make very clear. Indeed, the algorithmical and practical aspects of
% the phase retrieval problem from noisy data still pose significant
% challenges.

\subsection{Main approaches to phase retrieval}

Holographic techniques are among the more popular methods that have
been proposed to measure the phase of the optical wave. While
holographic techniques have been successfully applied in certain areas
of optical imaging, they are generally difficult to implement in
practice % and come with a number of challenges 
\cite{Dua11}.
%\ejc{EJC: Should we mention other approaches
% for measuring phase as that of W.~E.~Moerner \& Co?}  \ts{TS: Checked
%Moerner, he attempts high and superresolution, but does not deal
%with phase problem. So no need to cite him.}
Hence, the development of algorithms for signal recovery from magnitude
measurements is still a very active field of research.  Existing
methods for phase retrieval rely on all kinds of a priori information
about the signal, such as positivity, atomicity, support constraints,
real-valuedness, and so on \cite{Fie78,Fie82,Mar07,CMW07}.  Direct methods
\cite{Hau86} are limited in their applicability to small-scale
problems due to their large computational complexity. 
% \ejc{EJC: We need to be careful about the
%   last sentence. We might need to substantiate. References?}  \ts{TS:
% I would move the last two sentences starting with ``In truth ...", to the
% very end of this subsection, because
% we discuss another important method in the next paragraph. Therefore these
% two sentences, which are summarizing the situation appear to early. I have
% added more specific criticism about the alternating projection methods after
% we state Fienup's method. We already give concrete refs. for the drawbacks
% of holography and of direct methods. So when we move the two sentences 
% to the end of this subsection, we have already substantiated them 
% with individual criticism and references.}

Oversampling in the Fourier domain has been proposed as a means to
mitigate the non-uniqueness of the phase retrieval problem. While
oversampling offers no benefit for most one-dimensional signals, the
situation is more favorable for multidimensional signals, where it has
been shown that twofold oversampling in each dimension almost always
yields uniqueness for finitely supported, real-valued and non-negative
signals~\cite{BS79,Hay82,San85}. In other words, a digital image of
the form $x = \{x[t_1, t_2]\}$ with $0 \le t_1 < n_1$ and $0 \le t_2 <
n_2$, whose Fourier transform is given by
\begin{equation}
  \label{eq:Fourier2D}
  {\hat x}[\omega_1,\omega_2] = \frac{1}{\sqrt{n_1 n_2}} \sum x[t_1,
  t_2] e^{-i 2\pi (\omega_1 t_1/n_1 + \omega_2 t_2/n_2)}, 
\end{equation}
is usually uniquely determined from the values of $| {\hat
  x}[\omega_1,\omega_2]|$ on the oversampled grid $\omega = (\omega_1,
\omega_2) \in \Omega = \Omega_1 \times \Omega_2$ in which $\Omega_i =
\{0,1/2,1,3/2,\ldots, n_i + 1/2\}$. This holds provided $x$ has proper
spatial support, is real valued and non-negative.

As pointed out in \cite{LBL02}, these uniqueness results do not say
anything about how a signal can be recovered from its intensity
measurements, or about the robustness and stability of commonly used
reconstruction algorithms --- a fact we shall make very clear in the
sequel. In fact, theoretical uniqueness conditions do not readily
translate into numerical methods and as a result, the algorithmical
and practical aspects of the phase retrieval problem (from noisy data)
still pose significant challenges.

By and large, the most popular methods for phase retrieval from
oversampled data are alternating projection algorithms pioneered by
Gerchberg and Saxton \cite{GS72} and Fienup \cite{Fie78,Fie82}.  These
methods often require careful exploitation of signal constraints and
delicate parameter selection to increase the likelihood of convergence
to a correct solution \cite{NPC03,Mar07,CMW07,Mar07a}. We describe the
simplest realization of a widely used approach based on alternating
projections \cite{MKS00}, which assumes support constraints in the
spatial domain and oversampled measurements in the frequency domain.
With $T$ being a {\em known} subset containing the support of the
signal $x$ ($\text{supp}(x) \subset T$) and Fourier magnitude
measurements $\{y[\omega]\}_{\omega \in \Omega}$ with $y[\omega] =
|\hat{x}[\omega]|$, the method works as follows:
\begin{enumerate}
\item {\bf Initialization:} Choose an initial guess $x_0$ and set
  $z_0[\omega] = y[\omega]
  \frac{\hat{x}_0[\omega]}{|\hat{x}_0[\omega]|}$ for
  $\omega \in \Omega$. 
\item {\bf Loop:} For $k=1,2,\dots$ inductively define 
$$ (1) \qquad x_k[t] = 
\begin{cases}
(F^{-1} z_{k-1})[t] & \text{if $t \in T$,} \\
0               & \text{else;}
\end{cases} \hspace*{52mm}
$$
$$(2) \qquad
z_{k}[\omega] = y[\omega] \frac{\hat{x}_k[\omega]}{|\hat{x}_k[\omega]|} 
\qquad \,\text{for $\omega \in \Omega$}
\hspace*{50mm}$$
until convergence.
\end{enumerate}

While this algorithm is simple to implement and amenable to additional
constraints such as the positivity of $x$, its convergence remains
problematic.  Projection algorithms onto convex sets are well
understood \cite{Bre65,GPR67,You87,BCL02}.  However, the set $\{z :
|\hat{z}[\omega]| = |\hat{x}[\omega]|\}$ is not convex and, therefore,
the algorithm is not known to converge in general or even to give a
reasonable solution \cite{LH87,BCL02,LBL02}.  Good results have been
reported in certain settings but they appear to be nevertheless
somewhat problematic in light of our numerical experiments from
Section~\ref{sec:numerics}.  Moreover, as discussed
in~\cite{Pfeiffer}, one of the most stringent limitations of these
methods is the need for isolated objects (the support constraint).
Finally, \cite{Mar08} points out that oversampling is not always
practically feasible as certain experimental geometries allow only for
sub-Nyquist sampling; an example is the Bragg sampling from periodic
crystalline structures.

In a different direction, a frame-theoretic approach to phase
retrieval has been proposed in \cite{BCE06,BBC09}, where the authors
derive various necessary and sufficient conditions for the uniqueness
of the solution, as well as various numerical algorithms.  While
theoretically appealing, the practical applicability of these results
is limited by the fact that very specific types of measurements are
required, which cannot be realized in most applications of interest.

To summarize our discussion, we have seen many methods which all
represent some important attempts to find efficient algorithms, and
work well in certain situations. However, these techniques do not
always provide a consistent and robust result. % for a number of cases
% although successful instances have been reported

\subsection{PhaseLift -- a novel methodology}

This paper develops a novel methodology for phase retrieval based on a
rigorous and flexible numerical framework.  Whereas most of the
existing methods seek to overcome nonuniqueness by imposing additional
constraints on the signal, we pursue a different direction by assuming
no constraints at all on the signal. There are two main components to
our approach.
\begin{itemize}
\item {\em Multiple structured illuminations.}  We suggest collecting
  several diffraction patterns providing `different views' of the
  sample or specimen. This can be accomplished in a number of ways:
  for instance, by modulating the light beam falling onto the sample
  or by placing a mask right after the sample, see Section
  \ref{sec:method} for details. Taking multiple diffraction patterns
  usually yields uniqueness as discussed in Section \ref{sec:theory}.

  The concept of using multiple measurements as an attempt to resolve
  the phase ambiguity for diffraction imaging is of course not new,
  and was suggested in \cite{Mis73}. Since then, a variety of methods
  have been proposed to carry out these multiple measurements;
  depending on the particular application, these may include the use
  of various gratings and/or of masks, the rotation of the axial
  position of the sample, and the use of defocusing implemented in a
  spatial light modulator, see \cite{Dua11} for details and
  references. Other approaches include {\em ptychography}, an exciting
  field of research, where one records several diffraction patterns
  from overlapping areas of the sample, see \cite{Rod08,TDB09} and
  references therein.

\item {\em Formulation of phase recovery as a matrix completion
    problem.} We suggest (1) lifting up the problem of recovering a
  vector from quadratic constraints into that of a recovering of a
  rank-one matrix from affine constraints, and (2) relaxing the
  combinatorial problem into a convenient convex program.  
Since the lifting step is fundamental to our approach, we will 
refer to the proposed numerical framework as {\em PhaseLift}.
  The price we pay for trading the nonconvex quadratic constraints into 
  convex constraints is that we must deal with a highly underdetermined
  problem. However, recent advances in the areas of compressive
  sensing and matrix completion have shown that such convex
  approximations are often exact.

  Although our algorithmic framework appears to be novel for phase
  retrieval, the idea of solving problems involving nonconvex
  quadratic constraints by semidefinite relaxations has a long history
  in optimization, see \cite{BenTalNem} and references therein, and
  Section \ref{sec:precedents} below for more discussion.
\end{itemize}
The goal of this paper is to demonstrate that taken together, multiple
coded illuminations and convex programming (trace-norm minimization)
provide a powerful new approach to phase retrieval. Further, a
significant aspect of our methodology is that our systematic
optimization framework offers a principled way of dealing with noise,
and makes it easy to handle various statistical noise models. This is
important because in practice, measurements are always noisy.  In
fact, our framework can be understood as an elaborate regularized
maximum likelihood method. Lastly, our framework can also include
a priori knowledge about the signal that can be formulated or relaxed
as convex constraints.

\subsection{Precedents}
\label{sec:precedents}

At the abstract level, the phase retrieval problem is that of finding
$x \in \C^n$ obeying quadratic equations of the form $|\<a_k, x\>|^2 =
b_k$. Casting such quadratic constraints as affine constraints about
the matrix variable $X = x x^\star$ has been widely used in the
optimization literature for finding good bounds on a number of
quadratically constrained quadratic problems (QCQP). Indeed, solving
the general case of a QCQP is known to be an NP-hard problem since it
includes the family of boolean linear programs \cite{BenTalNem}.  The
approach usually consists in finding a relaxation of the QCQP using
semidefinite programming (SDP), for instance via Lagrangian
duality. An important example of this strategy is Max Cut, an NP-hard
problem in graph theory which can be formulated as a QCQP.  In a
celebrated paper, Goemans and Williamson introduced a relaxation
\cite{GoemansWilliamson} for this problem, which lifts or linearizes a
nonlinear, nonconvex problem to the space of symmetric
matrices. Although there are evident connections to our work, our
relaxation is quite different from these now standard techniques.

The idea of linearizing the phase retrieval problem by reformulating it as a
problem of recovering a matrix from linear measurements can be found
in \cite{BBC09}. While this reference also contains some intriguing
numerical recovery algorithms, their practical relevance for most
applications is limited by the fact that the proposed measurement
matrices either require a very specific algebraic structure which does not
seem to be compatible with the physical properties of diffraction,
or the number of measurements is proportional to the square of the signal
dimension, which is not feasible in most applications.

In terms of framework, the closest approach is the paper \cite{CMP10},
in which the authors use a matrix completion approach for array
imaging from intensity measurements. Although this paper executes a
similar relaxation as ours, there are some differences. We present
a ``noise-aware'' framework, which makes it possible to account for a 
variety of noise models in a systematic way. Moreover, our emphasis is
on a novel combination of structured illuminations and convex programming, 
which seems to bear great potential.
%\ejc{[EJC:
%  What are the differences we would like to emphasize? That we have
%  noise-aware variants? What would we like to say about George's
%  paper?]}  At the risk of repeating ourselves, it is the novel
%combination of structured illumination and convex programming that
%seems to bear great potential.

% \ejc{EJC: Should we have a short section
%   announcing {\em some} theoretical results?}
% \ts{TS: Let's first write the theory section and return to this once it is
% written. Then I have a better feeling whether a short theory announcement
% section is warranted or not}.

\section{Methodology}
\label{sec:method}

\subsection{Structured illumination}

Suppose $x = \{x[t]\}$ is the object of interest ($t$ may be a one- or
multi-dimensional index).  In this paper, we shall discuss
illumination schemes collecting the diffraction pattern of the
modulated object $w[t] x[t]$, where the waveforms or patterns $w[t]$
may be selected by the user. There are many ways in which this can be
implemented in practice, and we discuss just a few of those.

\begin{figure}
\begin{center}
\includegraphics[width=120mm]{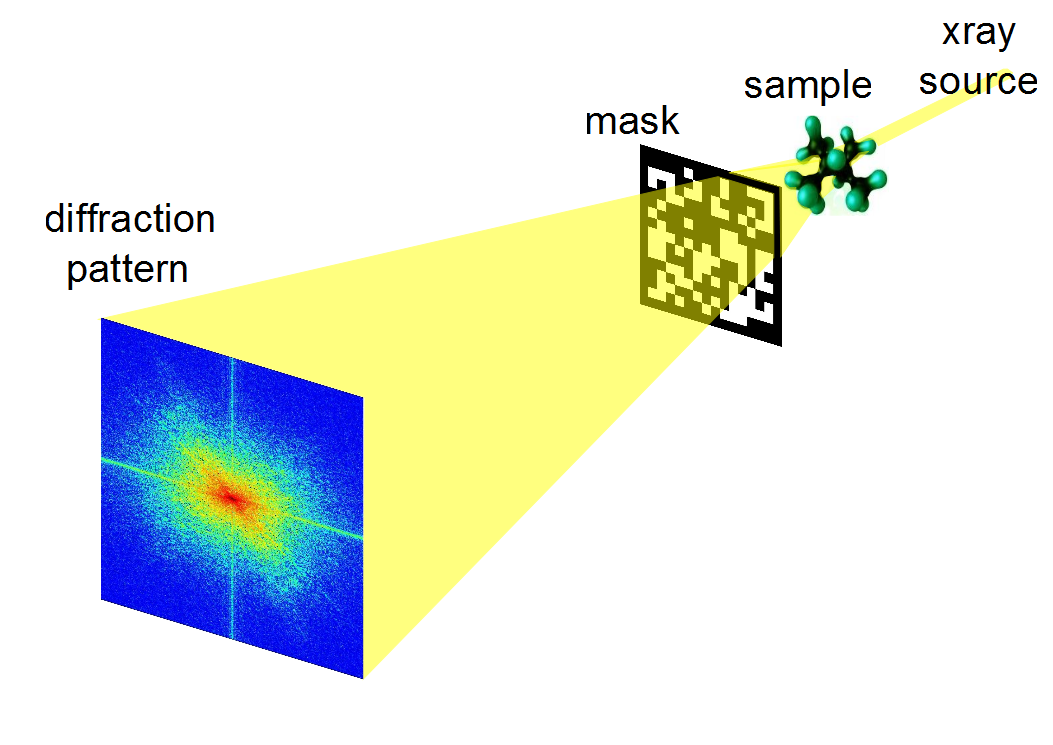}
\caption{A typical setup for structured illuminations in diffraction imaging
using a phase mask.}
\label{fig:mask}
\end{center}
\end{figure}

\begin{figure}
\begin{center}
\includegraphics[width=120mm]{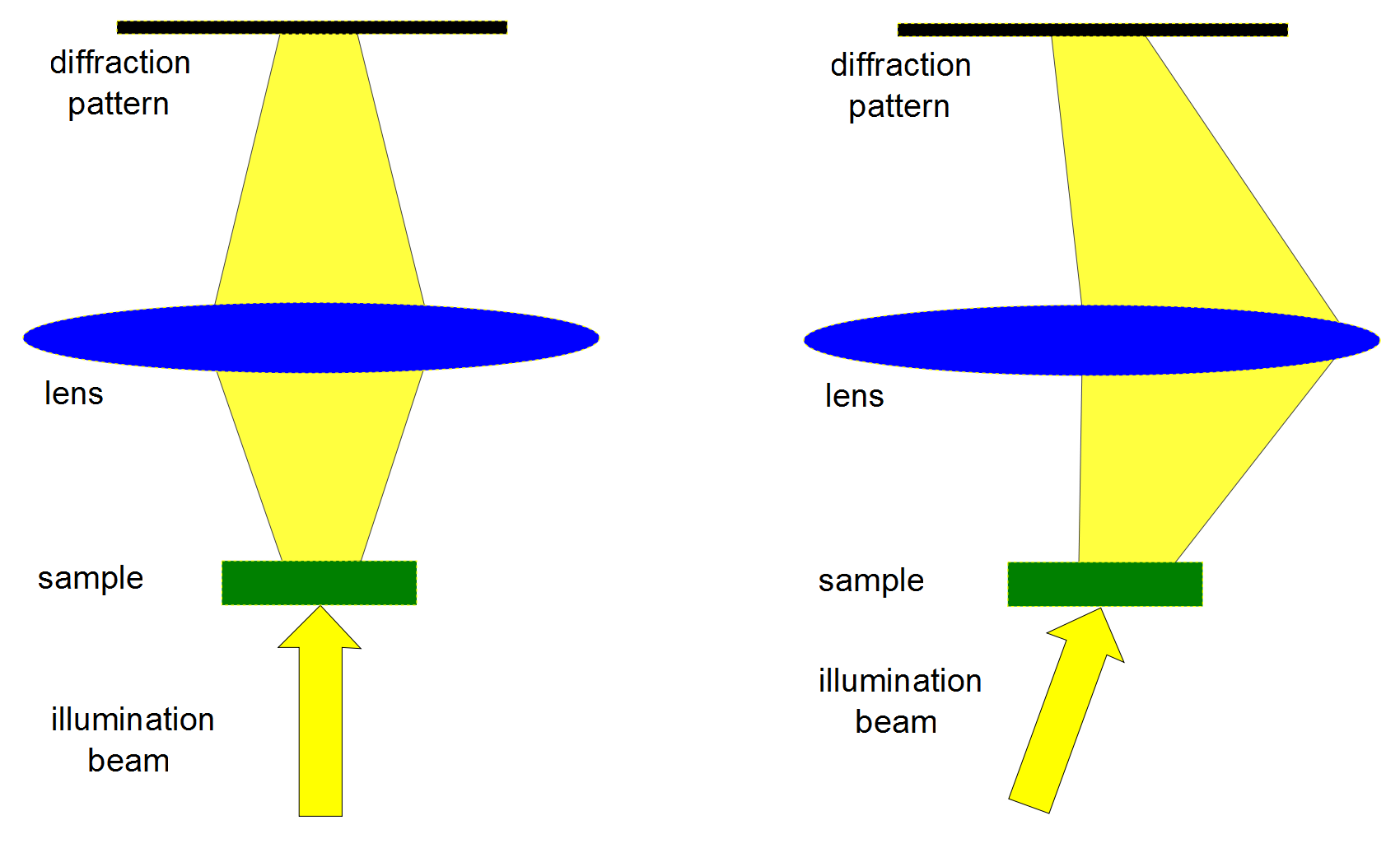}
\caption{A typical setup for structured illuminations in diffraction imaging
using oblique illuminations.  The
  left image shows direct (on-axis) illumination and the right image
  corresponds to oblique (off-axis) illumination.}
\label{fig:oblique}
\end{center}
\end{figure}

\begin{itemize}
\item {\em Masking.} One possibility is to modify the phase front
  after the sample by inserting a mask or a phase plate, see
  \cite{Liu08} for example. A schematic layout is shown in Figure
  \ref{fig:mask}. In \cite{Pfeiffer}, the sample is scanned by
  shifting the phase plate as in ptychography (discussed below); the
  difference is that one scans the known phase plate rather than the
  object being imaged.
\item {\em Optical grating.} Another standard approach would be to change
  the profile or modulate the illuminating beam, which can easily be
  accomplished by the use of optical gratings~\cite{LP97}. A simplified 
  representation would look similar to the scheme depicted in Figure
  \ref{fig:mask}, with a grating instead of the mask (the grating could be
  placed before or after the sample).

\item {\em Ptychography.}  Here, one measures multiple diffraction
  patterns by scanning a finite illumination on an extended
  specimen~\cite{Rod08,TDB09}.
  In this setup, it is common to maintain a substantial overlap
  between adjacent illumination positions.

\item {\em Oblique illuminations.} One can use illuminating beams
  hitting the sample at user specified angle \cite{FHP10}, see Figure
  \ref{fig:oblique} for a schematic illustration of this approach. One
  can also imagine having multiple simultaneous oblique illuminations.
\end{itemize}

As is clear, there is no shortage of options and one might prefer
solutions which require generating as few diffraction patterns as
possible for stable recovery.

\subsection{Lifting}

Suppose we have $x_0 \in \C^n$ or $\C^{n_1 \times n_2}$ (or some
higher-dimensional version) about which we have quadratic measurements
of the form
\begin{equation}
  \mapA(x_0) = \{|\<a_k, x_0\>|^2 : k = 1, 2, \ldots, m\}.
\label{nonlinearmap}
\end{equation}
In the setting where we would collect the diffraction pattern of $w[t]
x_0[t]$ as discussed earlier, then the waveform $a_k[t]$ can be written as 
\[
a_k[t] \propto w[t] e^{i2\pi \, \<\omega_k, t\>};
\]
here, $\omega_k$ is a frequency value so that $a_k[t]$ is a patterned
complex sinusoid. One can assume for convenience that the normalizing
constant is such that $a_k$ is unit normed, i.e.~$\|a_k\|_2^2 = \sum_t
|a_k[t]|^2 = 1$.  Phase retrieval is then the feasibility problem
\begin{equation}
\label{eq:retrieval}
  \begin{array}{ll}
    \text{find}   & \quad x\\ 
\text{obeying} & \quad  \mapA(x) = \mapA(x_0) := b.
  \end{array}
\end{equation}

As is well known, quadratic measurements can be lifted up and
interpreted as linear measurements about the rank-one matrix $X = x
x^*$. Indeed,
\[
|\<a_k, x\>|^2 = \trace(x^* a_k a_k^* x) = \trace(a_k a_k^* x x^*) :=
\trace(A_k X),
\]
where $A_k$ is the rank-one matrix $a_k a_k^*$. % For later reference
% it will be useful to denote 
% \begin{equation}
% A: = [a_1, a_2, \dots , a_m]^\ast.
% \label{framematrix}
% \end{equation}
In what follows, we
will let $\cA$ be the linear operator mapping positive semidefinite
matrices into $\{\trace(A_k X) : k = 1,\ldots, m\}$.  Hence, the phase
retrieval problem is equivalent to
\begin{equation}
\label{eq:rankmin}
  \begin{array}{ll}
    \text{find}   & \quad X\\ 
    \text{subject to} & \quad  \cA(X) = b\\
& \quad X \succeq 0\\
& \quad \rank(X) = 1
\end{array} 
\qquad \Leftrightarrow \qquad 
  \begin{array}{ll}
    \text{minimize}   & \quad \rank(X)\\ 
    \text{subject to} & \quad  \cA(X) = b\\
& \quad X \succeq 0. 
\end{array}
\end{equation}
Upon solving the left-hand side of \eqref{eq:rankmin}, we would
factorize the rank-one solution $X$ as $x x^*$, hence finding
solutions to the phase-retrieval problem. Note that the equivalence
between the left- and right-hand side of \eqref{eq:rankmin} is
straightforward since by definition, there exists a rank-one solution.
Therefore, our problem is a rank minimization problem over an affine
slice of the positive semidefinite cone. As such, it falls in the
realm of low-rank {\em matrix completion} or {\em matrix recovery}, a
class of optimization problems that has gained tremendous attention in
recent years, see e.g.~\cite{Recht07,CR08,CT09}. Just as in matrix
completion, the linear system $\cA(X) = b$, with unknown in the
positive semidefinite cone, is highly underdetermined. For instance
suppose our signal $x_0$ has $n$ complex unknowns. Then we may imagine
collecting six diffraction patterns with $n$ measurements for each (no
oversampling). Thus $m = 6n$ whereas the dimension of the space of $n
\times n$ Hermitian matrices over the reals is $n^2$, which is
obviously much larger.

We are of course interested in low-rank solutions and this makes the
search feasible. This also raises an important question: what is the
minimal number of diffraction patterns needed to recover $x$, whatever
$x$ may be? Since each pattern yields $n$ {\em real-valued}
coefficients and $x$ has $n$ {\em complex-valued} unknowns, the answer
is at least two. Further, in the context of quantum state tomography,
Theorem II in \cite{Fin04} shows one needs at least $3n-2$ intensity
measurements to guarantee uniqueness, hence suggesting an absolute
minimum of three diffraction patterns. Are three patterns sufficient?
For some answers to this question, see Section \ref{sec:theory}.

\subsection{Recovery via convex programming}

%\new{As indicated earlier, we will refer to the numerical framework
%which we introduce in this section as PhaseLift.}
The rank minimization problem \eqref{eq:rankmin} is NP hard. 
We propose using the trace norm as a convex surrogate
\cite{Beck98,Mesbahi97} for the rank functional, giving the familiar
SDP (and a crucial component of PhaseLift),
\begin{equation}
\label{eq:tracemin}
 \begin{array}{ll}
    \text{minimize}   & \quad \text{trace}(X)\\ 
    \text{subject to} & \quad  \cA(X) = b\\
& \quad X \succeq 0. 
\end{array}
\end{equation}
This problem is convex and there exists a wide array of numerical
solvers including the popular Nesterov's accelerated first order
method \cite{NesterovBook}. As far as the relationship between
\eqref{eq:rankmin} and \eqref{eq:tracemin} is concerned, the matrix
$\cA$ in most diffraction imaging applications is not known to obey
any of the conditions derived in the literature
\cite{CR08,CT09,Recht07} that would guarantee a formal equivalence
between the two programs. Nevertheless, the formulation
\eqref{eq:tracemin} enjoys great empirical performance as demonstrated
in Section \ref{sec:numerics}.

We mentioned earlier that measurements are typically noisy and that
our formulation allows for a principled approach to deal with this
issue for a variety of noise models. Suppose the measurement vector
$\{b_k\}$ is sampled from a probability distribution $p(\cdot ; \mu)$,
where $\mu = \mapA(x_0)$ is the vector of noiseless values, $\mu_k =
|\<a_k, x_0\>|^2$. Then a classical fitting approach simply consists
of maximizing the likelihood,
\begin{equation}
\label{eq:mle}
 \begin{array}{ll}
    \text{maximize}   & \quad p(b; \mu)\\ 
    \text{subject to} & \quad  \mu = \mapA(x)
\end{array}
\end{equation}
with optimization variables $\mu$ and $x$. (A more concise
  description is to find $x$ such that $ p(b; \mapA(x))$ is maximum.)
This is, of course, not tractable and our convex formulation suggests
solving instead
\begin{equation}
\label{eq:penalizedmle}
 \begin{array}{ll}
   \text{minimize}   & \quad -\log p(b; \mu) + \lambda \trace(X)\\ 
   \text{subject to} & \quad  \mu = \cA(X)\\
   & \quad X \succeq 0
\end{array}
\end{equation}
with optimization variables $\mu$ and $X$ (in other words, find $X
\succeq 0$ such that $-\log p(b; \cA(X)) + \lambda \trace(X)$ is
minimum). Above, $\lambda$ is a positive scalar and, hence, our
approach is a penalized or regularized maximum likelihood method,
which trades off between goodness and complexity of the fit. When the
likelihood is log-concave, problem \eqref{eq:penalizedmle} is convex
and solvable. We give two examples for concreteness:
\begin{itemize}
\item {\em Poisson data.} Suppose that $\{b_k\}$ is a sequence of
  independent samples from the Poisson distributions
  $\text{Poi}(\mu_k)$.  The Poisson log-likelihood for independent
  samples has the form $\sum_k b_k \log \mu_k - \mu_k$ (up to an
  additive constant factor) and thus, our problem becomes
\[
\begin{array}{ll}
  \text{minimize}   & \quad \sum_k  [\mu_k - b_k \log \mu_k] + \lambda \trace(X)\\ 
  \text{subject to} & \quad  \mu = \cA(X)\\
  & \quad X \succeq 0. 
\end{array}
\]
\item {\em Gaussian data.}  Suppose that $\{b_k\}$ is a sequence of
  independent samples from the Gaussian distribution with mean $\mu_k$
  and variance $\sigma_k^2$ (or is well approximated by Gaussian
  variables). Then our problem becomes
\[
\begin{array}{ll}
  \text{minimize}   & \quad \sum_k  \frac{1}{2\sigma^2_k} \, (b_k - \mu_k)^2 + 
  \lambda \trace(X)\\ 
  \text{subject to} & \quad  \mu = \cA(X)\\
  & \quad X \succeq 0. 
\end{array}
\]
If $\Sigma$ is a diagonal matrix with diagonal elements $\sigma_k^2$,
this can be written as
\[
\begin{array}{ll}
  \text{minimize}   & \quad \frac12 [b - \cA(X)]^* \Sigma^{-1} [b - \cA(X)]  + 
  \lambda \trace(X)\\ 
  \text{subject to} & \quad X \succeq 0. 
\end{array}
\]
\end{itemize}
Both formulations are of course convex and in both cases, one recovers
the noiseless trace-minimization problem \eqref{eq:tracemin} as
$\lambda \rightarrow 0^+$.

In addition, it is straightforward to include further constraints
frequently discussed in the phase retrieval literature such as
real-valuedness, positivity, atomicity and so on. Suppose the support
of $x$ is known to be included in a set $T$ known a priori. Then we
would add the linear constraint
\[
X_{ij} = 0, \quad (i,j) \notin T \times T. 
\]
(Algorithmically, one would simply work with a reduced-size matrix.)
Suppose we would like to enforce real-valuedness, then we simply
assume that $X$ is real valued and positive semidefinite. Finally
positivity can be expressed as linear inequalities
\[
X_{ij} \ge 0. 
\]
Of course, many other types of constraints can be incorporated in this
framework, which provides appreciable flexibility.

%\subsection{Improvement via reweighting}
\subsection{PhaseLift with reweighting}
\label{sec:reweight}

The trace norm promotes low-rank solutions and this is why it is often
used as a convex proxy for the rank. However, it is possible to
further promote low-rank solutions by solving a sequence
of weighted trace-norm problems, a technique which has been shown to
provide even more accurate solutions \cite{FHB03,CWB07}. The
reweighting scheme works like this: choose $\epsilon > 0$; start with
$W_0 = I$ and for $k = 0, 1, \ldots,$ inductively define $X_k$ as the
optimal solution to
\begin{equation}
\label{eq:reweight}
 \begin{array}{ll}
    \text{minimize}   & \quad \text{trace}(W_k X)\\ 
    \text{subject to} & \quad  \cA(X) = b\\
& \quad X \succeq 0 
\end{array}
\end{equation}
and update the `weight matrix' as
\[
W_{k+1} = (X_k + \epsilon I)^{-1}. 
\]
The algorithm terminates on convergence or when the iteration count
$k$ attains a specified maximum number of iterations
$k_{\text{max}}$. One can see that the first step of this procedure is
precisely \eqref{eq:tracemin}; after this initial step, the algorithm
proceeds in solving a sequence of trace-norm problems in which the
matrix weights $W_k$ are roughly the inverse of the current guess.

As explained in the literature \cite{FHB03,Faz02}, this
reweighting scheme can be viewed as attempting to solve 
\begin{equation}
\label{eq:logdet}
 \begin{array}{ll}
   \text{minimize}   & \quad f(X) =  \log(\det(X + \epsilon I)) \\ 
   \text{subject to} & \quad  \cA(X) = b\\
   & \quad X \succeq 0 
\end{array}
\end{equation}
% over the same set $\{X: \cA(X) = b \text{ and } X \succeq 0\}$ as in
% \eqref{eq:rankmin}--\eqref{eq:tracemin} 
by minimizing the tangent approximation to $f$ at each iterate; that is
to say, at step $k$, \eqref{eq:reweight} is equivalent to minimizing
$f(X_{k-1}) + \<\nabla f(X_{k-1}), X - X_{k-1}\>$ over the feasible
set. This can also be applied to noise-aware variants where one would
simply replace the objective functional in \eqref{eq:penalizedmle}
with
\[
-\log p(b; \mu) + \lambda \trace(W_k X),
\]
at each step, and update $W_k$ in exactly the same way as before.

The log-det functional is closer to the rank functional than the trace
norm. In fact, minimizing this functional solves the phase retrieval
problem as incorporated in the following theorem.
\begin{theorem}
\label{teo:reweight}
Suppose $\mapA$ is one to one and that the identity matrix $I$ is in
the span of the sensing matrices $A_k$.
% there exists scalars $h_1,
% \ldots, h_m$ obeying $\sum_k h_k a_k a_k^* = \sum_k h_k A_k = I$.
Then the unique solution of the phase retrieval problem
\eqref{eq:retrieval} is also the unique minimizer to
\eqref{eq:logdet}---up to global phase.\footnote{Quadratic
  measurements can of course never distinguish between $x$ and $c x$
  in which $c \in \C$ has unit norm. When the solution is unique up to
  a multiplication with such a scalar, we say that unicity holds up to
  global phase. From now on, whenever we talk about unicity, it is
  implied up to global phase.} This holds for all values of $\epsilon
> 0$. The same conclusion holds without the inclusion assumption
provided one modifies the reweighting scheme and substitutes the
objective function $f(X)$ in \eqref{eq:logdet} with $f(R X R)$ where
$R = (\sum_k A_k)^{1/2}$.
\end{theorem}
Since the reweighting algorithm is a good heuristic for solving
\eqref{eq:logdet}, we potentially have an interesting and tractable
method for phase retrieval. It is not a perfect heuristic, however, as
we cannot expect this procedure to always find the global minimum
since the objective functional is concave.

The assumption that the identity matrix is in the span of the $A_k$'s
holds whenever the modulus of the Fourier transform of the sample is
measured.  Indeed, if $|F x|^2$ is observed, then letting $\{f_k^*\}$
be the rows of $F$, we have $\sum_k f_k f_k^* = I$.

\begin{proof} Our assumption implies that for any feasible $X$,
  $\trace(X) = \sum_k h_k \trace(A_k X) = \sum_k h_k b_k$ is
  fixed. Assume without loss of generality that feasible points obey
  $\trace(X) = 1$ (if $\trace(X) = 0$, then the unique solution is $X
  = 0$). If $x_0$ is the unique solution to phase retrieval (up to
  global phase), then $X_0 = x_0 x_0^*$ is the only rank-one feasible
  point. We thus need to show that any feasible $X$ with $\rank(X) >
  1$, obeys $f(X) > f(X_0)$, a fact which follows from the strong
  concavity of $f$ (of the logarithm). Let $X = \sum_j \lambda_j u_j
  u_j^*$ be any eigenvalue decomposition of a feasible point. Then
\[
f(X) = \log(\det(\epsilon I + X)) = \sum_j \log(\epsilon + \lambda_j),
\]
and it follows from the strict concavity of the log that 
\[
\sum_j \log(\epsilon + \lambda_j) > \sum_j \lambda_j \log(\epsilon +
1) + (1-\lambda_j) \log \epsilon = \log(\epsilon + 1) + (n-1) \log
\epsilon.
\]
The first strict inequality holds unless $X$ is rank one, in which
case, we have equality. The equality follows from $\sum_j \lambda_j =
\trace(X) = 1$. Since the right-hand side is nothing else than
$f(X_0)$, the theorem is established.

For the second part, set $B_k = R^{-1} A_k R^{-1}$ and consider a new
data problem with constraints $\{X: \trace(B_k X) = b_k, X \succeq
0\}$. Now $X$ is feasible for our problem if and only if $RXR$ is
feasible for this new problem. (This is because the mapping $X \mapsto
R X R$ preserves the positive semidefinite cone.) Now suppose $x_0$ is
the solution to phase retrieval and set $X_0 = x_0 x_0^*$ as
before. Since $I$ is in the span of the sensing matrices $B_k$, we
have just learned that for all for all $RXR \neq RX_0R$ and $X$
feasible for our problem,
\[
f(R X R) > f(R X_0 R).
\]
This concludes the proof.
\end{proof}

\section{Theory}
\label{sec:theory}

Our PhaseLift framework poses two main theoretical questions:
\begin{enumerate}
\item When do multiple diffracted images imply unicity of the solution?
\item When does our convex heuristic succeed in recovering the unique
  solution to the phase-retrieval problem?
\end{enumerate}
Developing comprehensive answers to these questions constitutes a
whole research program, which clearly is beyond the scope of this
work. In this paper, we shall limit ourselves to introducing some
theoretical results showing simple ways of designing diffraction
patterns, which give unicity. Our focus is on getting uniqueness from
a very limited number of diffraction patterns.  For example, we shall
demonstrate that in some cases three diffraction images are
sufficient for perfect recovery. Thus, we give below partial answers
to the first question and will address the second in a later
publication.

A frequently discussed approach to retrieve phase information uses a
technique from holography. Roughly speaking, the idea is to let the
signal of interest $x$ interfere with a known reference beam $y$. One
typically measures $|x + y|^2$ and $|x - iy|^2$ and precise knowledge
of $y$ allows, in principle, to recover the amplitude and phase of
$x$. Holographic techniques are hard to implement \cite{Dua11} in
practice. Instead, we propose using a modulated version of the signal
itself as a reference beam which in some cases may be easier to
implement.

To discuss this idea, we need to introduce some notation. For a
complex signal $z \in \C^n$, we let $|z|^2$ be the nonnegative
real-valued $n$-dimensional vector containing the squared magnitudes
of $z$.  Suppose first that $x$ is a one-dimensional signal $(x[0],
x[1], \ldots, x[n-1])$ and that $F_n$ is the $n \times n$ unitary
DFT. In this section, we consider taking $3n$ real-valued measurements
of the form
\begin{equation}
  \label{eq:data}
  \mapA(x) = \{|F_n x|^2, |F_n (x + D^s x)|^2, |F_n(x - iD^s x)|^2\}, 
\end{equation}
where $D$ is the modulation 
\[
D = \text{diag}(\{e^{i2\pi t/n}\}_{0 \le t \le n-1}).
\]
These measurements can be obtained by illuminating the sample with the
three light fields $1$, $e^{i2\pi st/n}$ and $e^{i2\pi (st/n -
  1/4)}$. We show below that these $3n$ measurements are generally
sufficient for perfect recovery.

\begin{theorem}
\label{teo:1D}
Suppose the DFT of $x \in C^n$ does not vanish. Then $x$ can be
recovered up to global phase from the $3n$ real numbers $\mapA(x)$
\eqref{eq:data} if and only if $s$ is prime with $n$. In particular,
assuming primality, if the trace-minimization program
\eqref{eq:tracemin} or the iteratively reweighted algorithm return a
rank-1 solution, then this solution is exact.

Conversely, if the DFT vanishes at two frequency points $k$ and $k'$
obeying $k - k' \neq s \text{ mod } n$, then recovery is not possible.
from the $3n$ real numbers \eqref{eq:data}.
\end{theorem}
The proof of this theorem is constructive and we give a simple
algorithm that achieves perfect reconstruction. Further, one can use
masks to scramble the Fourier transform as to make sure it does not
vanish. Suppose for instance that we collect
\[
\mapA(Wx), \quad W = \text{diag}(\{z[t]\}_{0 \le t \le n-1}).
\]
where the $z[t]$'s are iid $\mathcal{N}(0,1)$. Then since the Fourier
transform of $z[t] x[t]$ does not vanish with probability one, we have
the following corollary. 

\begin{corollary}
  Assume $s$ is prime with $n$.  Then with probability one, $x$ can be
  recovered up to global phase from the $3n$ real numbers $\mapA(Wx)$
  where $W$ is the diagonal matrix with Gaussian entries above.
\end{corollary}
Of course, one could derive similar results by scrambling the Fourier
transform with the aid of other types of masks, e.g.~binary masks. We
do not pursue such calculations. 

\newcommand{\cF}{\mathcal{F}}

We now turn our attention to the situation in higher dimensions and
will consider the 2D case (higher dimensions are treated in the same
way). Here, we have a discrete signal $x[t_1, t_2] \in \C^{n_1 \times
  n_2}$ about which we take the $3n_1 n_2$ measurements
\begin{equation}
  \label{eq:data2D}
  \{|\cF_{n_1 \times n_2}  x|^2, |\cF_{n_1 \times n_2} (x + \mathcal{D}^s x)|^2, |\cF_{n_1 \times n_2} (x - i\mathcal{D}^s x)|^2\}, \quad s = (s_1, s_2); 
\end{equation}
$\cF_{n_1 \times n_2}$ is the 2D unitary Fourier transform defined by
\eqref{eq:Fourier2D} in which the frequencies belong to the 2D grid
$\{0, 1, \ldots, n_1 -1\} \times \{0, 1, \ldots, n_2 -1\}$, and
$\mathcal{D}^s$ is the modulation
$$
[\mathcal{D}^s x][t_1, t_2] = e^{i2\pi s_1 t_1/n_1} \, e^{i2\pi s_2
  t_2/n_1} x[t_1, t_2].
$$
With these definitions, we have the following result: 
\begin{theorem}
  \label{teo:2D}
  Suppose the DFT of $x \in C^{n_1 \times n_2}$ does not vanish. Then
  $x$ can be recovered up to global phase from the $3n_1 n_2$ real
  numbers \eqref{eq:data2D} if and only if $s_1$ is prime with $n_1$,
  $s_2$ is prime with $n_2$ and $n_1$ is prime with $n_2$. Under these
  assumptions, if the trace-minimization program \eqref{eq:tracemin}
  or the iteratively reweighted algorithm return a rank-1 solution,
  then this solution is exact.
\end{theorem}

Again, one can apply a random mask to turn this statement into a
probabilistic statement holding either with probability one or with
very large probability depending upon the mask that is used. 

One can always choose $s_1$ and $s_2$ such that they be prime with
$n_1$ and $n_2$ respectively. The last condition may be less friendly
but one can decide to pad one dimension with zeros to guarantee
primality. This is equivalent to a slight oversampling of the DFT
along one direction. An alternative is to take $5 n_1 n_2$
measurements in which we modulate the signal horizontally and then
vertically; that is to say, we modulate with $s = (s_1,0)$ and then
with $s = (0,s_2)$. These $5 n_1 n_2$ measurements guarantee recovery
if $s_1$ is prime with $n_1$ and $s_2$ is prime with $n_2$ for all
sizes $n_1$ and $n_2$, see Section \ref{sec:ext} for details.

\subsection{Proof of Theorem \ref{teo:1D}}

Let $\hat{x} = (\hat x [0], \ldots, \hat x [n-1])$ be the DFT of $x$.
Then knowledge of $\mapA(x)$ is equivalent to knowledge of
\[
|\hat{x}[k]|^2, |\hat x[k] + \hat{x}[k-s]|^2, \text{ and } |\hat x[k]
-i \hat{x}[k-s]|^2
\]
for all $k \in \{0,1,\ldots, n-1\}$ (above, $k-s$ is understood mod
$n$). Write $\hat x[k] = |\hat x [k]| e^{i\phi[k]}$ so that $\phi[k]$
is the missing phase, and observe that
\begin{align*}
  |\hat x[k] + \hat{x}[k-s]|^2 & = |\hat x[k]|^2 + |\hat{x}[k-s]|^2 +
  2|\hat x[k]| |\hat x[k-s]| \text{Re}(e^{i(\phi[k-s] - \phi[k]})\\
  |\hat x[k] -i\hat{x}[k-s]|^2 & = |\hat x[k]|^2 + |\hat{x}[k-s]|^2 +
  2|\hat x[k]| |\hat x[k-s]| \text{Im}(e^{i(\phi[k-s] - \phi[k]}).
\end{align*}
Hence, if $\hat{x}[k] \neq 0$ for all $k \in \{0,1, \ldots, n-1\}$,
our data gives us knowledge of all phase shifts of the form
\[
\phi[k-s] - \phi[k], \quad k = 0,1, \ldots, n-1.
\]
We can, therefore, initialize $\phi[0]$ to be zero and then get the
values of $\phi[-s]$, $\phi[-2s]$ and so on.

This process can be represented as a cycle in the group $\Z/n\Z$ as
the sequence $(0,-s,-2s,\ldots)$. We would like this cycle to contain
$n$ unique elements, which is true if and only if the cyclic subgroup
$(0, s, 2s, \ldots )$ has order $n$. This is equivalent to requiring
$\text{gcd}(s,n) = 1$.  If this subgroup has a smaller order, then
recovery is impossible since we finish the cycle before we have all
the phases; the phases that we are able to recover do not enable us to
determine any more phases without making further assumptions.

For the second part of the theorem, assume without loss of generality,
that $s = -1$ and that $(k,k') = (0,k_0)$ ($1 < k_0 < n-1$). For
simplicity suppose these are the only zeros of the DFT. This creates
two disjoint sets of frequency indices: those for which $0 < k < k_0$
and those for which $k_0 < k \le n-1$. We are given no information
about the phase difference between elements of these two subgroups,
and hence recovery is not possible.  This argument extends to
situations where the DFT vanishes more often, in which case, we have
even more indeterminacy.

\subsection{Proof of Theorem \ref{teo:2D}}

Let $\hat{x} = \{\hat x [k_1,k_2]\}$, where $(k_1, k_2) \in \{0, 1,
\ldots n_1 - 1\} \times \{0, 1, \ldots, n_2-1\}$ be the DFT of $x$.
Then we have knowledge of 
\[
|\hat{x}[k_1,k_2]|^2, |\hat x[k_1,k_2] + \hat{x}[k_1-s_s, k_2 -
s_2]|^2, \text{ and } |\hat x[k_1,k_2] -i \hat{x}[k_1-s_1, k_2-s_2]|^2
\]
for all $(k_1 k_2)$. With the same notations as before, this gives 
knowledge of all phase shifts of the form
\[
\phi[k_1-s_1, k_2 - s_2] - \phi[k_1,k_2], \quad 0 \le k_1 \le n_1,
0\le k_2 \le n_2 -1. 
\]
Hence, we can initialize $\phi[0,0]$ to be zero and then get the
values of $\phi[-s_1, -s_2]$, $\phi[-2s_1, -2s_2]$ and so on. The
argument is as before: we would like the cyclic subgroup $\bigl(
(0,0), (s_1, s_2), (2s_1, 2s_2), \ldots \bigr)$ in $\Z/n_1 \Z \times
\Z/n_2 \Z$ to have order $n_1 n_2$. Now the order of an element $(s_1,
s_2) \in \Z/n_1 \Z \times \Z/n_2 \Z$ is equal to
\[
\text{lcm}(|s_1|, |s_2|) = \text{lcm}(n_1/\text{gcd}(n_1,s_1),
n_2/\text{gcd}(n_2,s_2)), 
\]
where $|s_1|$ is the order of $s_1$ in $\Z/n_1 \Z$ and likewise for
$|s_2|$.  Noting that $\text{lcm}(a, b) \le ab$ and that equality is
achieved if and only if $\text{gcd}(a, b) = 1$, we must simultaneously
have 
\[
\text{gcd}(s_1, n_1) = 1, \quad \text{gcd}(s_2, n_2) = 1 \quad \text{and }
\text{gcd}(n_1,n_2)) = 1
\]
to have uniqueness.

\subsection{Extensions}
\label{sec:ext}

It is clear from our analysis that if we were to collect
$|\cF_{n_1\times n_2} x|^2$ together with
\[
\{|\cF_{n_1 \times n_2} (x + \mathcal{D}^{s_k} x)|^2, |\cF_{n_1 \times
  n_2} (x - i\mathcal{D}^{s_k} x)|^2\}, \quad k = 1, \ldots, K, 
\]
so that one collects $(2K+1)n_1 n_2$ measurements, then 2D recovery is
possible if and only if $\{s_1, \ldots, s_K\}$ generates $\Z/n_1 \Z
\times \Z/n_2 \Z$ (and the Fourier transform has no nonzero
components). This can be understood by analyzing the generators of the
group $\Z/n_1 \Z \times \Z/n_2 \Z$.

A simple instance consists in choosing one modulation pattern to be
$(s_1, 0)$ and another to be $(0,s_2)$. If $s_1$ is prime is $n_1$ and
$s_2$ with $n_2$, these two modulations generate the whole group
regardless of the relationship between $n_1$ and $n_2$. An algorithmic
way to see this is as follows. Initialize $\phi(0,0)$. Then by using
horizontal modulations, one recovers all phases of the form
$\phi(k_1,0)$. Further, by using vertical modulations (starting with
$\phi(k_1,0)$), one can recover all phases of the form $\phi(k_1,k_2)$
by moving upward.

\section{Numerical Experiments}
\label{sec:numerics}

This section introduces numerical simulations to illustrate and study
the effectiveness of PhaseLift. % our methodology.

\subsection{Numerical solvers}
 
All numerical algorithms were implemented in Matlab using
TFOCS~\cite{BCG10} as well as modifications of TFOCS template
files. TFOCS is a library of Matlab-files designed to facilitate the
construction of first-order methods for a variety of convex
optimization problems, which include those we consider. 

In a nutshell, suppose we wish to solve the problem
\begin{equation}
\label{eq:composite}
 \begin{array}{ll}
   \text{minimize}   & \quad g(X) :=  -\ell(b; \cA(X)) + \lambda \trace(X) \\
   \text{subject to}    & \quad X \succeq 0 
\end{array}
\end{equation}
in which $\ell(b; \cA(X))$ is a smooth and concave (in $X$)
log-likelihood. Then a projected gradient method would start with an
initial guess $X_0$, and inductively define 
\[
X_k = \mathcal{P}(X_{k-1} - t_k \nabla g(X_{k-1})),  
\]
where $\{t_k\}$ is a sequence of stepsize rules and $\mathcal{P}$ is
the projection onto the positive semidefinite cone.  (Various stepsize
rules are typically considered including fixed stepsizes, backtracking
line search, exact line search and so on.)

TFOCS implements a variety of accelerated first-ordered methods
pioneered by Nesterov, see \cite{NesterovBook} and references therein.
One variant \cite{FISTA} works as follows. Choose $X_0$, set $Y_0 =
X_0$ and $\theta_0 = 1$, and inductively define
\begin{align*}
  X_k & = \mathcal{P}(Y_{k-1} - t_k \nabla g(Y_{k-1})) \cr
  \theta_k & = 2\Bigl[1 + \sqrt{1 + 4/\theta_{k-1}^2}\Bigr]^{-1} \cr
  \beta_k  & = \theta_k(\theta_{k-1}^{-1} - 1) \cr
  Y_k  & = X_k + \beta_k (X_k - X_{k-1})
\end{align*}
where $\{t_k\}$ is a sequence of stepsize rules as before. The
sequence $\{\theta_k\}$ is usually referred to as a sequence of
accelerated parameters, and $\{Y_k\}$ is an auxiliary sequence at
which the gradient is to be evaluated.  The advantage of this approach
is that the computational work per iteration is as in the projected
gradient method but the number of iterations needed to reach a certain
accuracy is usually much lower \cite{NesterovBook}. TFOCS implements
such iterations and others like it but with various improvements.

For large problems, e.g.~images with a large number $N$ of pixels, it
is costly to hold the $N \times N$ optimization variable $X$ in
memory. To overcome this issue, our computational approach maintains a
low-rank factorization of $X$. This is achieved by substituting the
projection onto the semidefinite cone (the expensive step) with a
proxy. Whereas $\mathcal{P}$ dumps the negative eigenvalues as in
\[
\mathcal{P}(X) = \sum_i \max(\lambda_i,0) u_i u_i^*,
\]
where $\sum_i \lambda_i u_i u_i^*$ ($\lambda_1 \ge \lambda_2 \ge
\ldots \ge \lambda_N$) is any eigenvalue decomposition of $X$, our
proxy only keeps the $k$ largest eigenvalues in the expansion as in
\[
\sum_{i \le k} \max(\lambda_i,0) u_i u_i^*.
\]
For small values of $k$---we use $k$ between 10 and 20---this can be
efficiently computed since we only need to compute the top
eigenvectors of a low-rank matrix at each step. Although this
approximation gives good empirical results, convergence is no longer
guaranteed.

\subsection{Setup}

To measure performance, we will use the mean-square error
(MSE). However, since a solution $x_0$ is only unique up to global
phase, it makes no sense to compute the squared distance between $x_0$
and the recovery $\hat{x}_0$.  Rather, we compute the distance to the
solution space, i.e.~we are interested in the relative MSE
defined as 
\[
\min_{c : |c| = 1} \,\, \frac{\|c x_0 - \hat{x}_0\|_2^2}{\|x_0\|^2}.
\]
This is the definition we will adopt throughout the
paper;\footnote{Alternatively, we could use $\|x_0 x_0^* - \hat{x}_0
  \hat{x}_0^*\|_F/\|x_0 x_0^*\|_F$, which gives very similar values.}
the Signal-to-Noise Ratio (SNR) is analogous, namely, $\text{SNR} =
\log_{10} (\text{rel.~MSE})$.

Although our algorithm favors low-rank solutions, it is not guaranteed
to find a rank-one solution.  Therefore, if our optimal solution
$\hat{X}_0$ does not have exactly rank one, we extract the rank-one
approximation $\hat{x}_0 \hat{x}_0^*$ where $\hat{x}_0$ is an
eigenvector associated with the largest eigenvalue. We use a scaling
such that $\|\hat{x}_0\|^2 = \|x_0\|^2$. Note that the $\ell_2$ norm
of the true solution is generally known since by Parseval's theorem,
the $\ell_2$ norm of $F x_0$ is equal to $\|x_0\|$. Hence, observing
the diffraction pattern of the object $x_0$ reveals its squared
$\ell_2$ norm.

\subsection{1-D simulations}  \label{ss:num1d}

Phase retrieval for one-dimensional signals arises in fiber optics
\cite{CM97,YM99,ILB08}, speech recognition \cite{RJ93}, 
but also in the determination of concentration profiles in diffraction
imaging~\cite{Bun07}. We evaluate  % the proposed algorithm 
PhaseLift for noiseless and
noisy data using a variety of different `illuminations' and test signals.

\subsubsection{Noisefree measurements} \label{sss:1dnoisefree}

\begin{figure}
\begin{center}
\subfigure[Smooth signal and its reconstruction]{
\includegraphics[width=70mm]{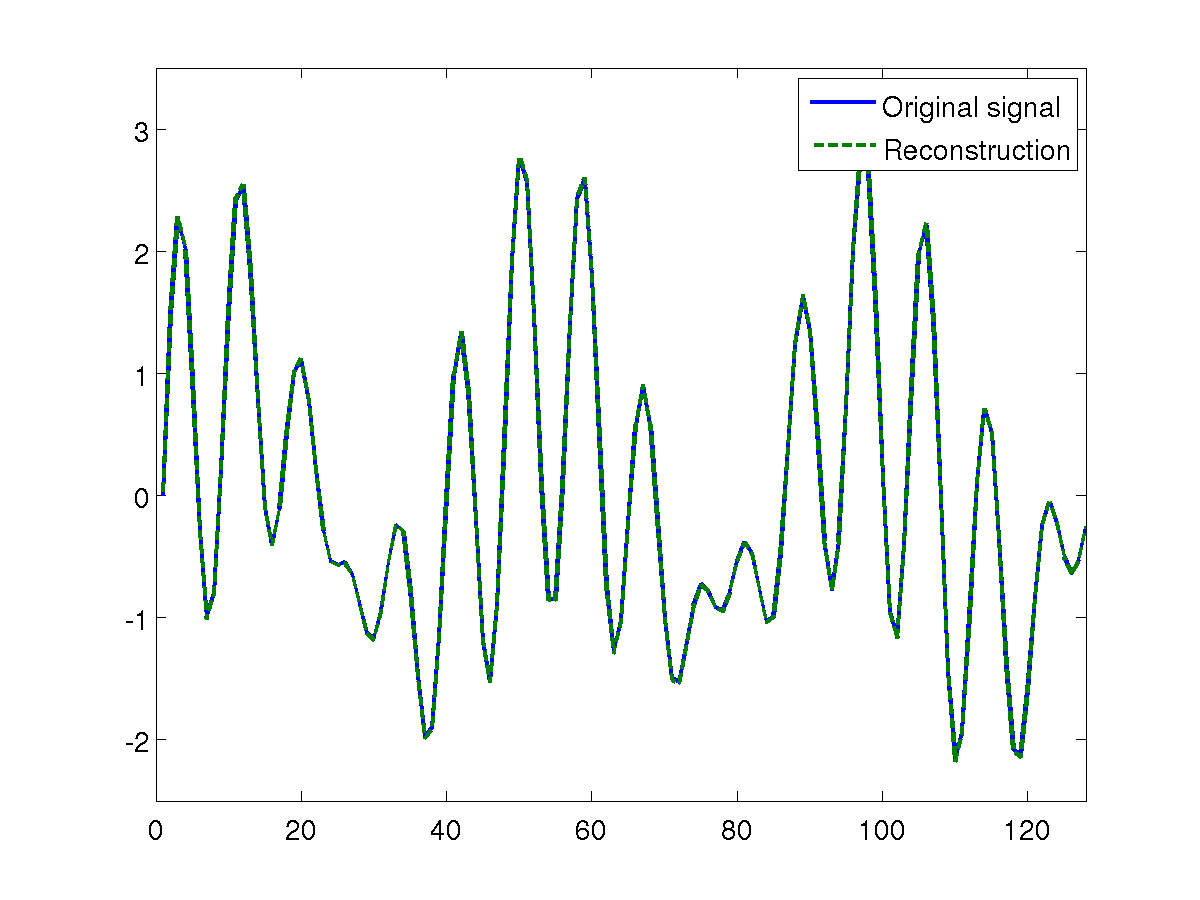}}
\qquad
\subfigure[Random signal and its reconstruction (real part only)]{
\includegraphics[width=70mm]{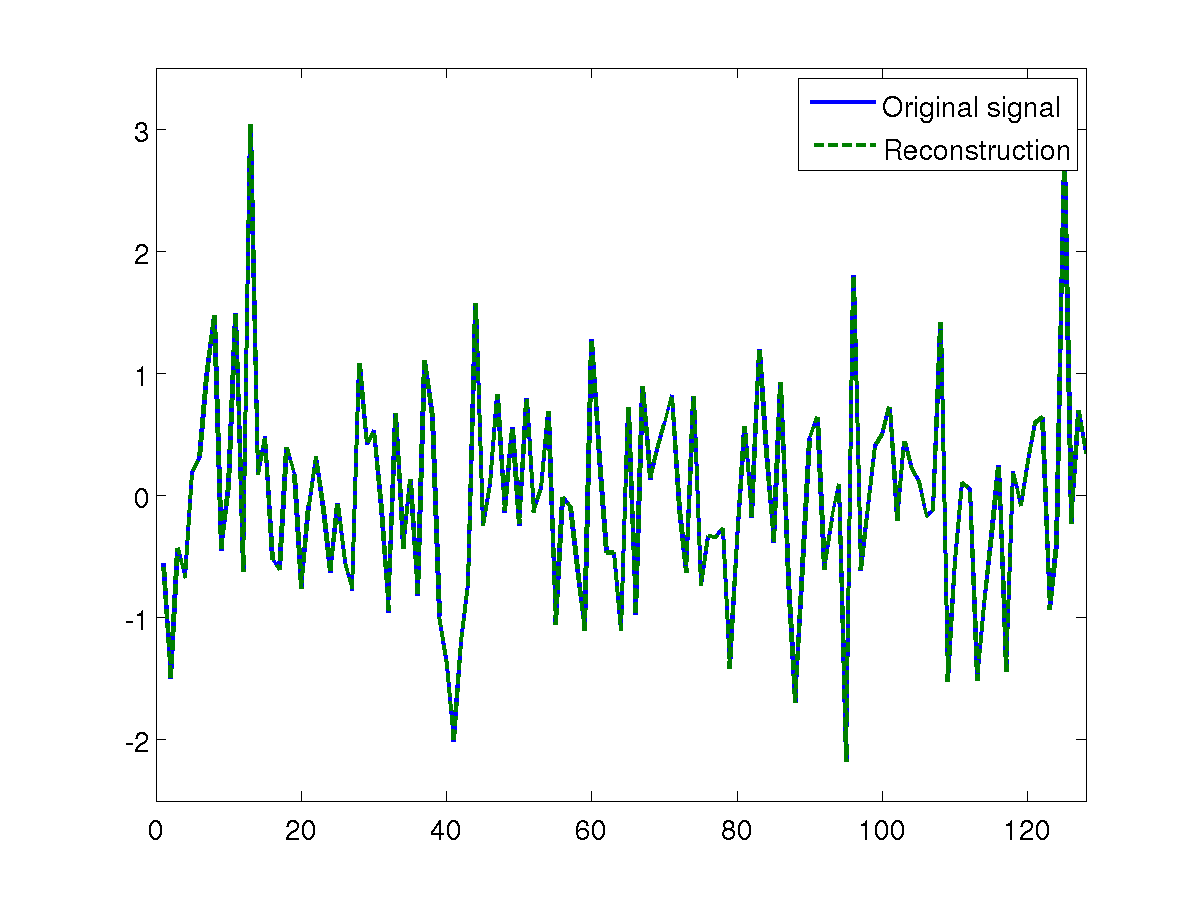}}
\caption{Two test signals and their reconstructions. The recovered
  signals are essentially indistinguishable from the originals.}
\label{fig:1A}
\end{center}
\end{figure}

In the first set of experiments we demonstrate the recovery of two very
different signals from noiseless data. Both test signals are of length
$n = 128$.  The first signal, shown in Figure \ref{fig:1A}(a)) is a
linear combination of a few sinusoids and represents a typical
transfer function one might encounter in optics. The second signal is
a complex signal, with independent Gaussian complex entries (each
entry is of the form $a + i b$ where $a$ and $b$ are independent
$\mathcal{N}(0,1)$ variables) so that the real and imaginary parts are
independent white noise sequences; the real part of the signal is
shown in Figure \ref{fig:1A}(b).

Four random binary masks are used to perform the structured
illumination so that we measure $|Ax|^2$ in which 
\[
A = F \begin{bmatrix} W_1 \\ W_2 \\ W_3 \\ W_4 \end{bmatrix}, 
\]
where each $W_i$ is diagonal with either $0$ or $1$ on the diagonal,
resulting in a total of 512 intensity measurements.  We work with the
objective functional $\frac12 \|b - \cA(X)\|_2^2 + \lambda \trace(X)$
and the constraint $X \succeq 0$ to recover the signal, in which we
use a small value for $\lambda$ such as $0.05$ since we are dealing
with noisefree data. We apply the reweighting scheme as discussed in
Section \ref{sec:reweight}.  (To achieve perfect reconstruction, one
would have to let $\lambda \to 0$ as the iteration count increases.)
The algorithm is terminated when the relative {\em residual error } is
less than a fixed tolerance, namely, $\|\cA(\hat{x}_0 \hat{x}_0^*) -
b\|_2 \le 10^{-6} \|b\|_2$, where $\hat{x}_0$ is the reconstructed
signal just as before.  The original and recovered signals are plotted
in Figure \ref{fig:1A}(a) and (b).  The SNR is 87.3dB in the first
case and 90.5dB in the second.

We have repeated these experiments with the same test signals and the
same algorithm, but using Gaussian masks instead of binary masks. In
other words, the $W_i$'s have Gaussian entries on the diagonal. It
turns out that in this case, three illuminations---instead of
four---were sufficient to obtain similar performance. This seems to be
empirical support for a long-standing conjecture in quantum mechanics
due to Ron Wright (see e.g.\ the concluding section
of~\cite{Vogt78}). The conjecture is that there exist three unitary
operators $U_1, U_2, U_3$ such that the phase of the
(finite-dimensional) signal $x$ is uniquely determined by the
measurements $|U_1 x|, |U_2 x|, |U_3 x|$. Our simulations suggest that
one can choose $U_1 = F$, $U_2 = F W$, and $U_3 = F W'$, where $W, W'$
are diagonal matrices with i.i.d.\ complex normal random variables as
diagonal entries. The choice $U_1 = F$, $U_2 = I$, $U_3 = F W$ was
equally successful in our experiments, and is a bit closer to the
quantum mechanical setting.  Furthermore, we point out that no
reweighting was needed, when we used six or more Gaussian
masks. Expressed differently, plain trace-norm minimization succeeds
with $6n$ or more intensity measurements of this kind.

\subsubsection{Noisy measurements} \label{sss:1dnoisy}

In the next set of experiments, we consider the case when the
measurements are contaminated with Poisson noise. The test signal is
again a complex random signal as above. Eight illuminations with
binary masks are used.  We add random Poisson noise to the
measurements for five different SNR levels, ranging from about 16dB to
about 52dB. Since the solution is known, we have calculated
reconstructions for various values of the parameter $\lambda$
balancing the negative log-likelihood and the trace norm, and report
results for that $\lambda$ giving the lowest MSE.  We implemented this
strategy via the standard Golden Section Search~\cite{Kie53}. In
practice one would have to find the best $\lambda$ via a strategy like
cross validation (CV) or generalized cross validation (GCV). For each
SNR level we repeated the experiment ten times with different random
noise and different binary masks.

\begin{figure}
\begin{center}
\includegraphics[width=100mm]{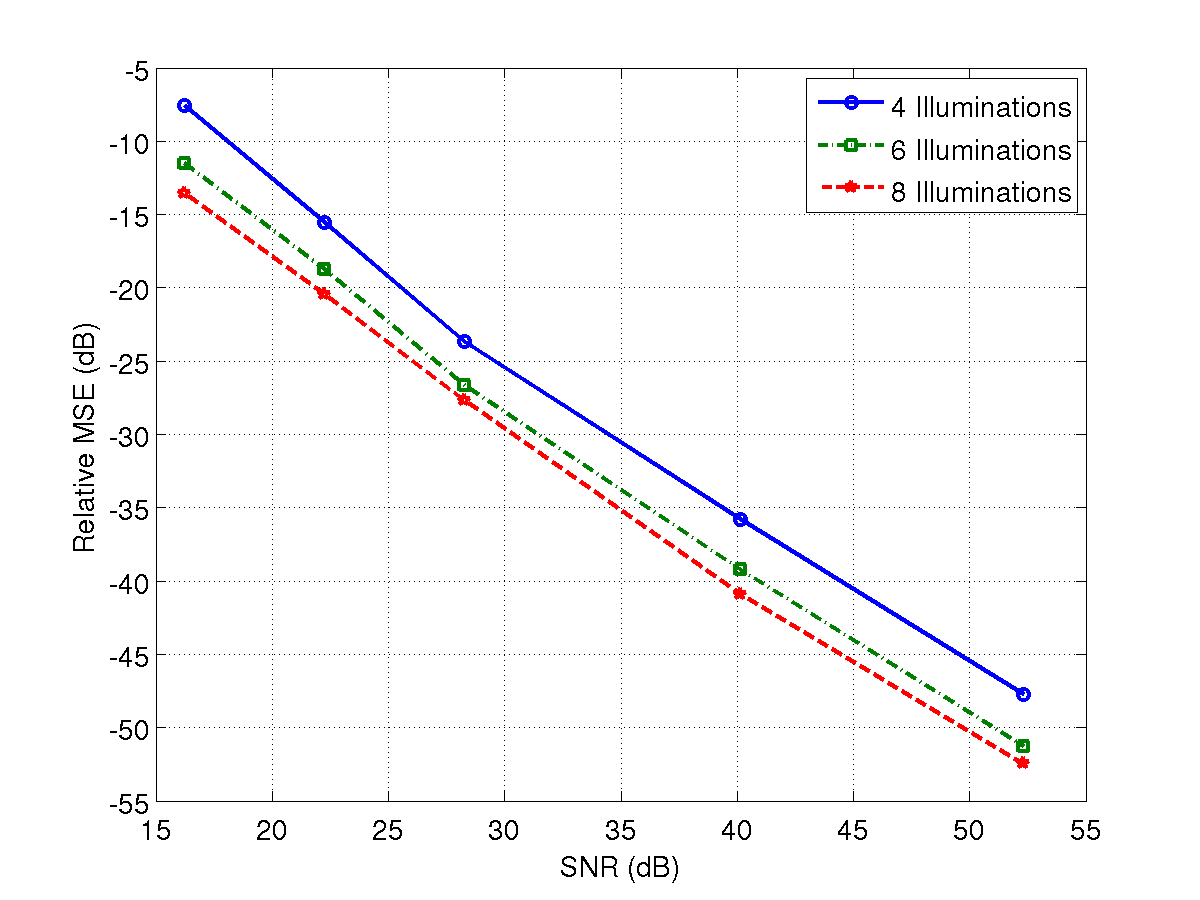}
\caption{SNR versus relative MSE on a dB-scale for different numbers of
illuminations with binary masks. The linear relationship between SNR and
MSE is apparent.}
\label{fig:1B}
\end{center}
\end{figure}

Figure~\ref{fig:1B} shows the average relative MSE in dB (the values
of $10 \log_{10}(\text{rel.~MSE})$ are plotted) versus the SNR. The
error curves show clearly the desirable linear behavior between SNR
and MSE with respect to the log-log scale. The performance degrades very
gracefully with decreasing SNR. Furthermore, the difference
of about 5dB between the error curve associated with four measurement
and the error curve associated with eight measurements corresponds to
an almost twofold error reduction, which is about as much improvement
as one can hope to gain by doubling the number of measurements.

\subsubsection{Multiple measurements via oversampling}
\label{sss:1doversampling}

It is well-known that for one-dimensional signals, oversampling does
not result in unique solutions to the phase problem \cite{LBL02}.
This might mainly be a theoretical issue without real practical
consequences.  Our numerical simulations confirm that this is not the
case and demonstrate very clearly that oversampling is not useful for
most one-dimensional signals.  We apply one of Fienup's algorithms as
discussed in Section 4.A in \cite{BCL02}, and %our algorithm 
PhaseLift
with reweighting to real-valued non-negative random signals of length
128. We use oversampling rates ranging from 2 to 5, and stop the
algorithms when the relative residual error is less than $10^{-3}$.

\begin{table}[h]
\centering
        \begin{tabular}{|c|c|c|c|c|}
          \hline
          Oversampling          & 2      & 3      & 4      & 5      \\
          \hline
          Relative MSE (Fienup) & 0.6811 & 0.6548 & 0.6180 & 0.6150 \\
          \hline
          Relative MSE (PhaseLift)   & 0.5451 & 0.5343 & 0.5299 & 0.4930 \\
          \hline
        \end{tabular}
        \caption{MSE obtained by Fienup's algorithm and by PhaseLift
          from oversampled DFT measurements. 
          Both methods produce guesses that fit the data and at 
          the same time, are 
          far from the true solution. This indicates that oversampling 
          the DFT results in an ill-posed problem since we have 
          several distinct solutions (probably infinitely many). }
\label{tab:1C}
\end{table}

The results are displayed in Table \ref{tab:1C}.  As can be seen, both
algorithms find nearly perfect fits to the data and yet, they are very
far from the original solution. Hence, we have a problem with multiple
solutions.  The average SNR of the ``reconstructions'' obtained via
Fienup's method is around 4dB, which is just barely better than if we
had used a random guess as a solution. The average SNR for the
solutions computed by PhaseLift %our algorithm 
is only marginally better.

\subsection{2-D simulations}  \label{ss:num2d}

We consider a stylized version of a setup one encounters in X-ray
crystallography or diffraction imaging.  The test image, shown in
Figure~\ref{fig:2A}(a) (magnitude), is a complex-valued image of size
$256 \times 256$, whose pixel values correspond to the complex
transmission coefficients of a collection of gold balls embedded in a
medium.

\subsubsection{Noisefree measurements} \label{sss:2dnoisefree}

\begin{figure}
\begin{center}
\subfigure[Original image]{
\includegraphics[width=60mm,height=60mm]{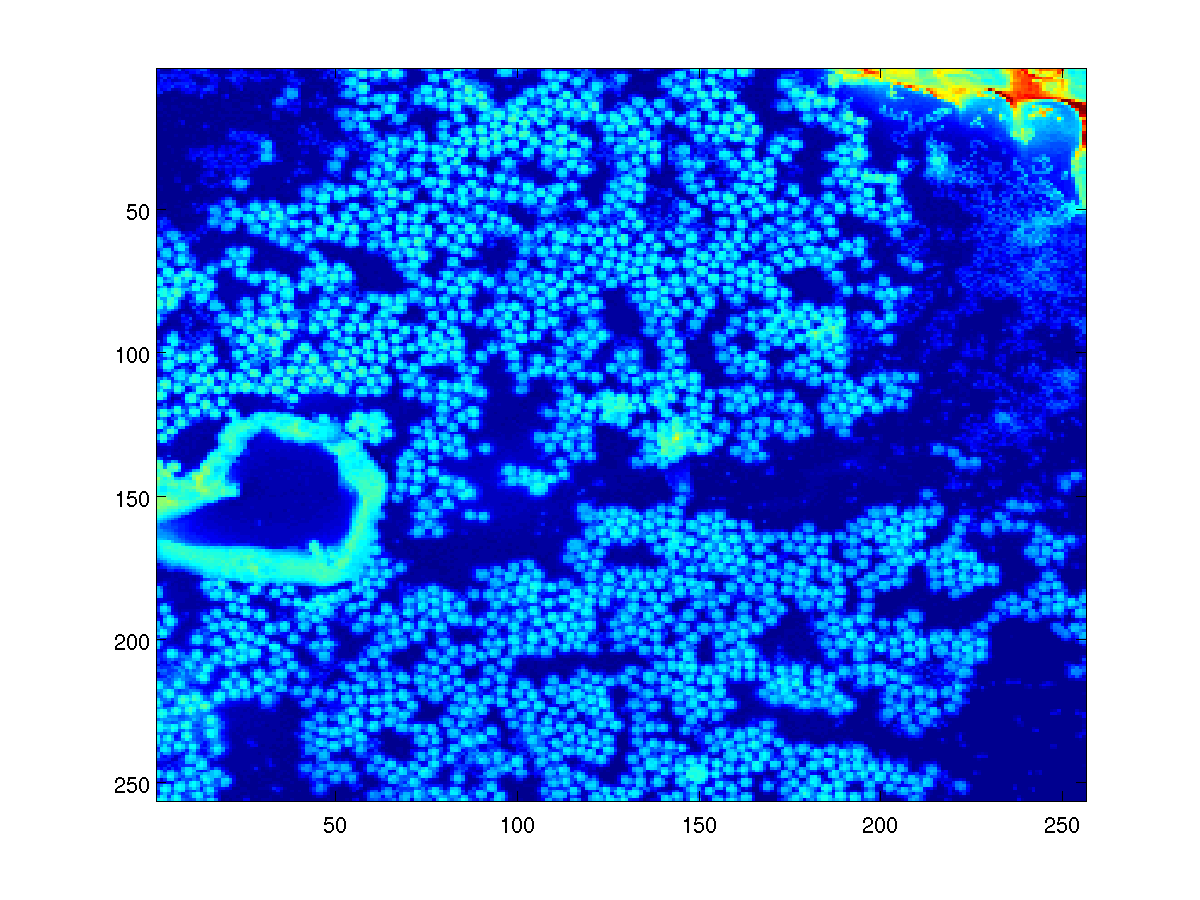}}
\quad
\subfigure[Reconstruction using 3 Gaussian masks]{
\includegraphics[width=60mm,height=60mm]{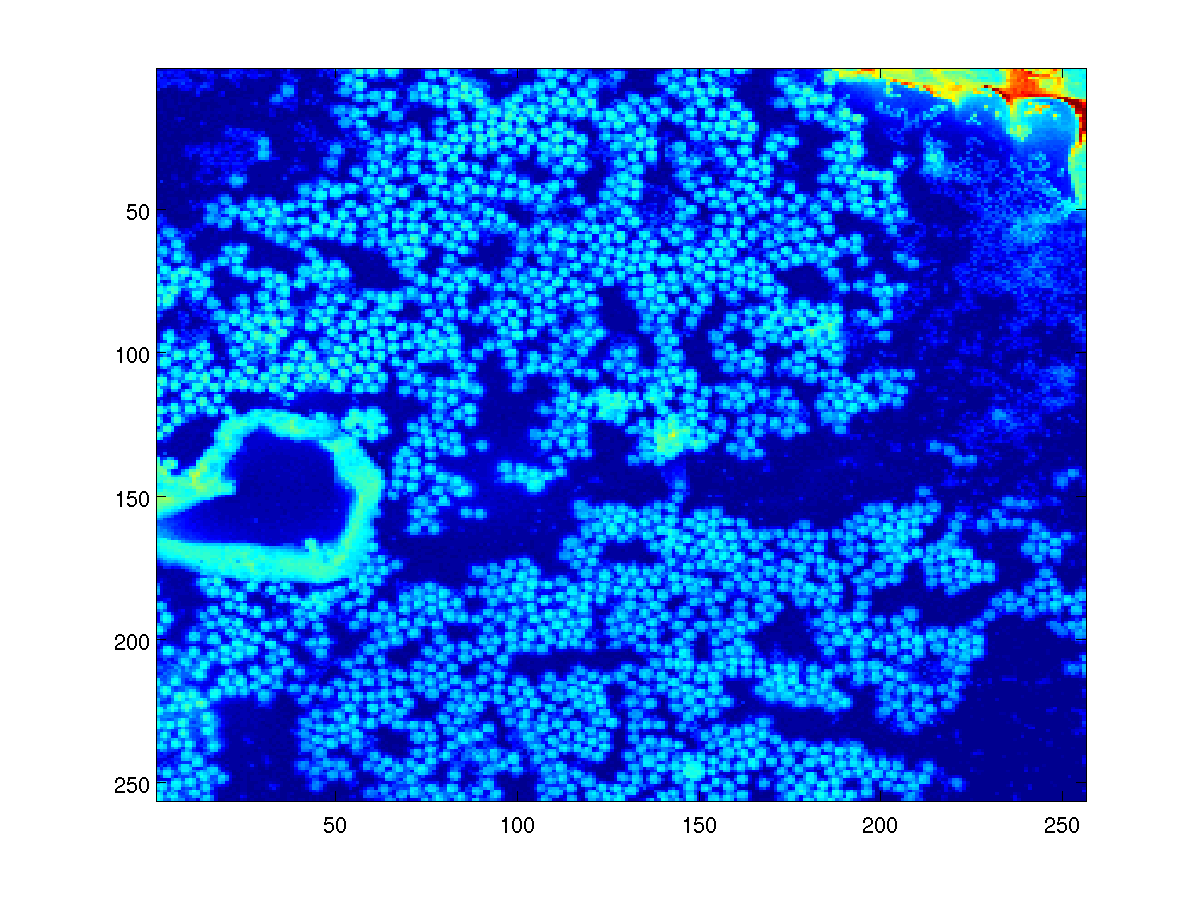}}
\subfigure[Reconstruction using 8 binary masks]{
  \includegraphics[width=60mm,height=60mm]{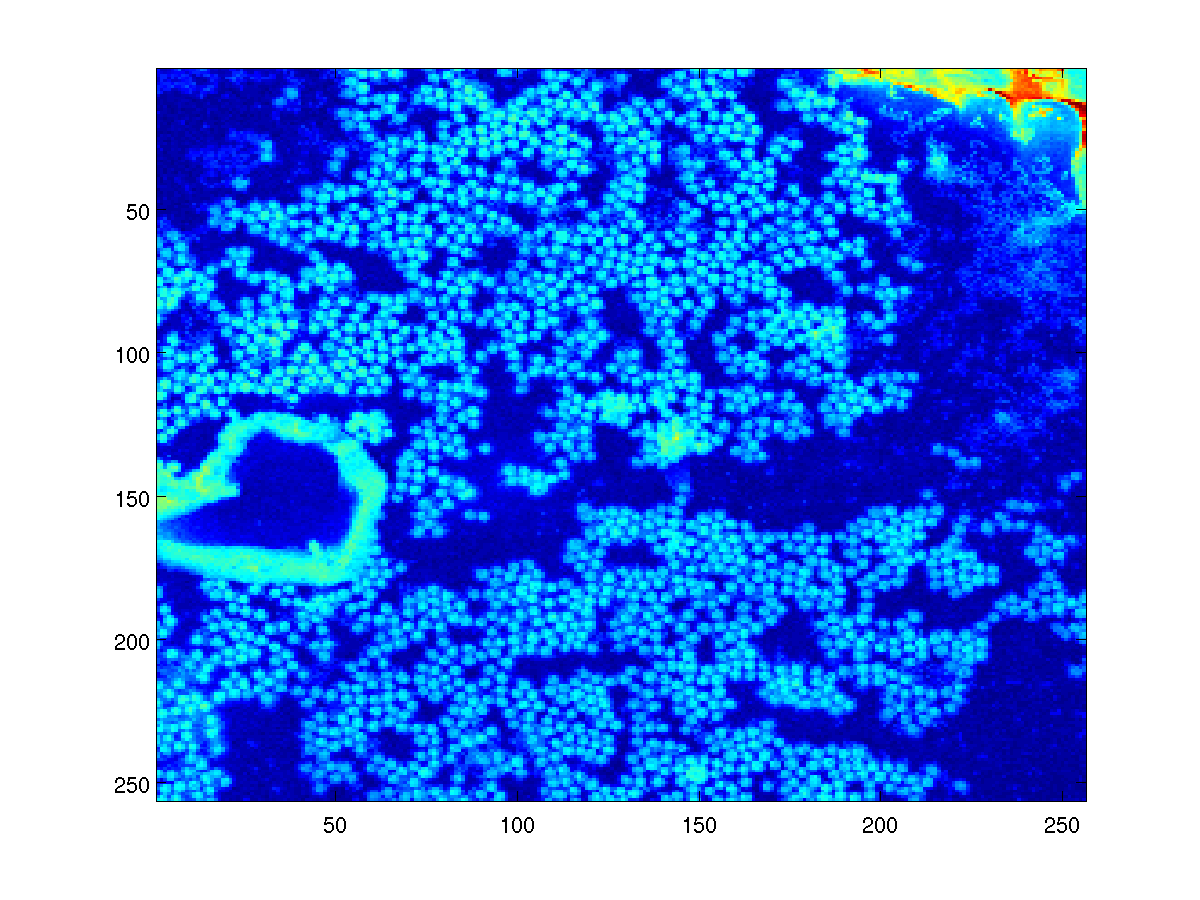}}
\quad \subfigure[Error between (a) and (c)]{
\includegraphics[width=60mm,height=60mm]{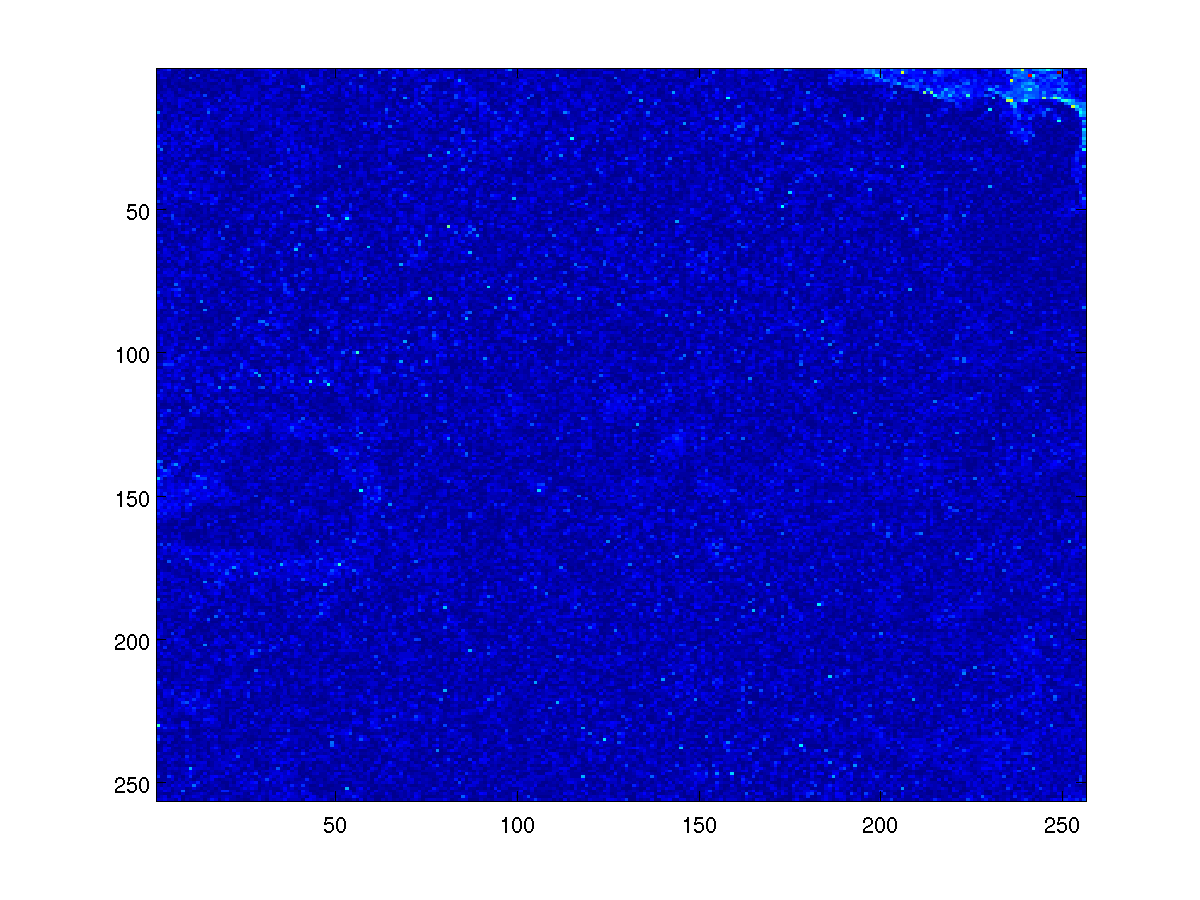}}
\caption{Original goldballs image and reconstructions via PhaseLift.}
\label{fig:2A}
\end{center}
\end{figure}

In the first experiment, we demonstrate the recovery of the image
shown in Figure~\ref{fig:2A}(a) from noiseless measurements. We
consider two different types of illuminations. The first uses Gaussian
random masks in which the coefficients on the diagonal of $W_k$ are
independent real-valued standard normal variables. 
We use three illuminations, one being constant, i.e.~$W_1
= I$, and the other two Gaussian. Again, we choose a small value of
$\lambda$ set to $0.05$ in $\frac12 \|b - \cA(X)\|_2^2 + \lambda
\trace(X)$ since we have no noise, and stop the reweighting iterations
as soon as the residual error obeys $\|\cA(\hat{x}_0 \hat{x}_0^*) -
b\|_2 \le 10^{-3} \|b\|_2$.  The reconstruction, shown in Figure
\ref{fig:2A}(b), is visually indistinguishable from the original.
Since the original image and the reconstruction are complex-valued,
we only display the absolute value of each image throughout this and
the next subsection.

Gaussian random masks may not be realizable in practice. Our second
example uses simple random binary masks, where the entries are either
$0$ or $1$ with equal probability.  In this case, a larger number of
illuminations as well as a larger number of reweighting steps are
required to achieve a reconstruction of comparable quality. The result
for eight binary illuminations is shown in Figure \ref{fig:2A}(c).

\subsubsection{Noisy measurements} \label{sss:2dnoisy}

\begin{figure}
\begin{center}
\subfigure[Low SNR]{
\includegraphics[width=60mm,height=60mm]{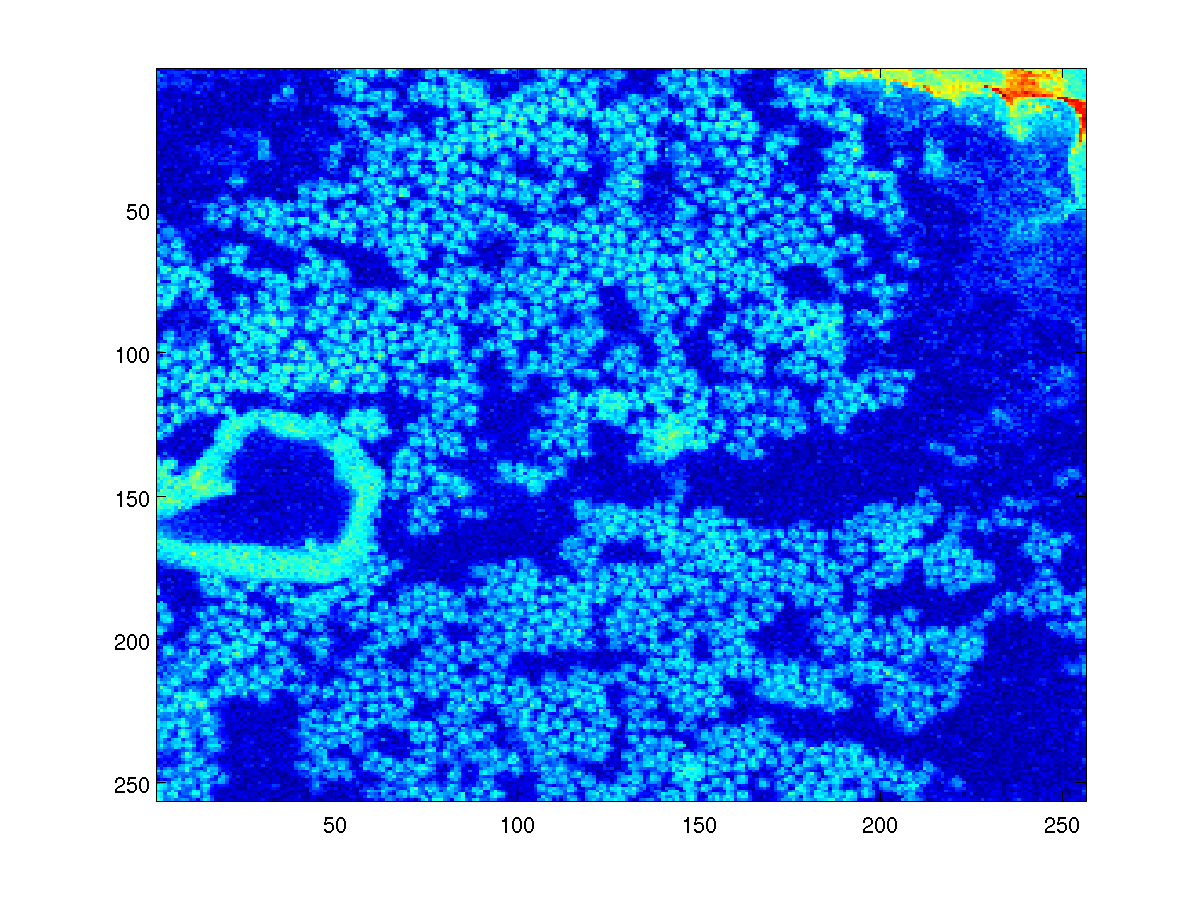}}
\qquad
\subfigure[High SNR]{
\includegraphics[width=60mm,height=60mm]{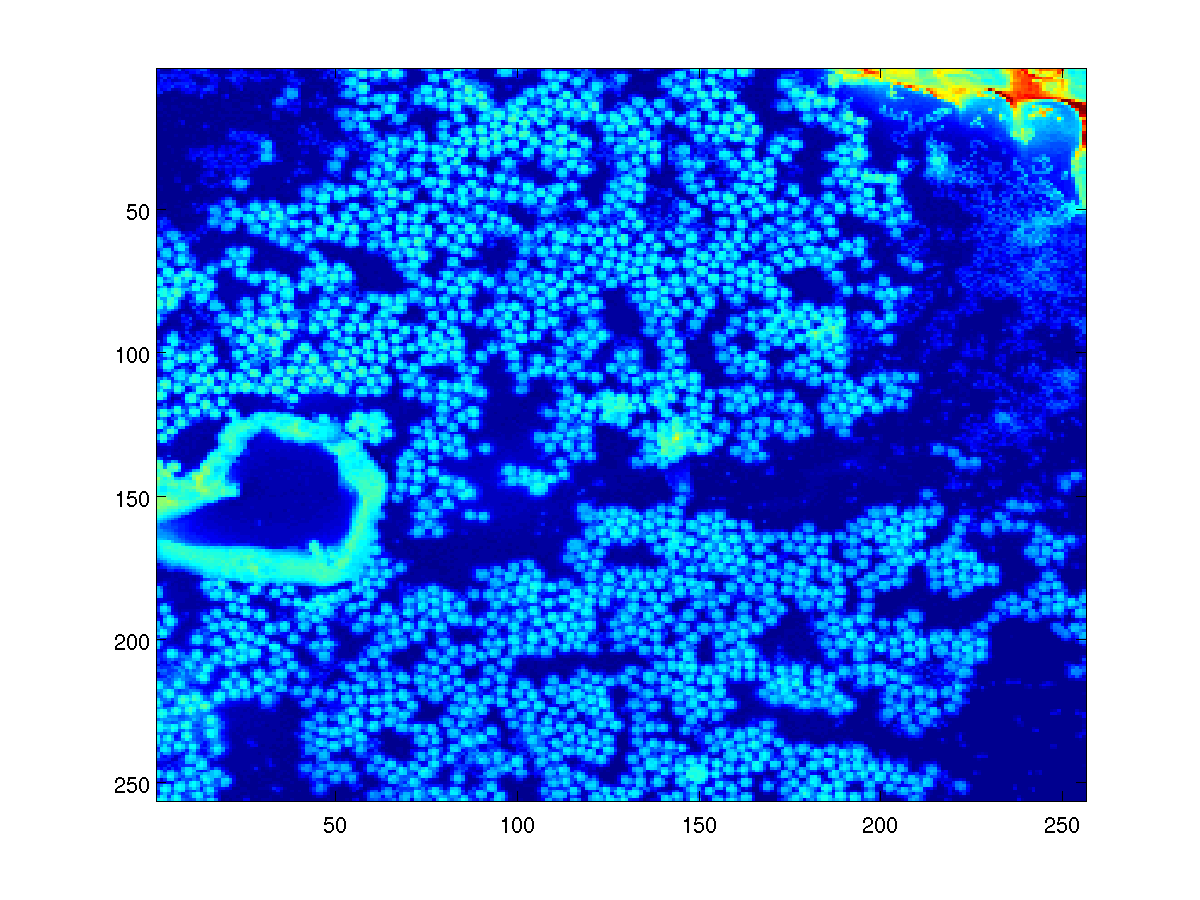}}
\caption{Reconstructions from noisy data via PhaseLift using 32 Gaussian random masks.}
\label{fig:2B}
\end{center}
\end{figure}

In the second set of experiments we consider the same test image as
before, but now with noisy measurements. In the first experiment the
SNR is 20dB, in the second experiment the SNR is 60dB.  We use 32
Gaussian random masks in each case. The resulting reconstructions are
depicted in Figure~\ref{fig:2B}(a) (20dB case) and
Figure~\ref{fig:2B}(b) (60dB case). The SNR in the 20dB case is
11.83dB. While the reconstructed image appears slightly more ``fuzzy''
than the original image, all features of the image are clearly
visible. In the 60dB case the SNR is 47.96dB, and the reconstruction is
virtually indistinguishable from the original image.

\subsubsection{Multiple measurements via oversampling}
\label{sss:2doversampling} 

Oversampling of two-dimensional signals is widely used to overcome the
nonuniqueness of the phase retrieval problem. We now explore whether
this approach is viable.

Here, we consider signals with real, non-negative values as test
images, a case frequently considered in the literature, see
e.g.~\cite{Mil90,MSC98,MKS00}. These images are of size $128 \times
128$.  We take noiseless measurements and apply PhaseLift % methodology 
as well as Fienup's iterative algorithms \cite[Section 4.A]{BCL02}. For
each method, we terminate the iterations if the relative residual
error is less than $10^{-3}$ or if the relative error between two
successive iterates is less than $10^{-6}$.  
\begin{itemize}
\item The simulations show that %our method 
PhaseLift yields reconstructions that
  fit the measured data well, yielding a small relative residual error
  $\|\cA(X) - y\|_2/\|y\|_2$, yet the reconstructions are far away
  from the true signal.  This behavior is indicative of an
  ill-conditioned problem.  
\item The iterates of Fienup's algorithm stagnated most of the time without
  converging to a solution. At other times it did yield
  reconstructions that fit the measured data well, but in either case
  the reconstruction was always very different from the true signal.
  Moreover, the reconstructions vary widely depending on the initial
  (random) guess.
\end{itemize}
Table~\ref{tab:2C} displays the results of PhaseLift %our algorithm 
as well as one of Fienup's algorithms, namely, the Error Reduction Algorithm
\cite[Section 4.~A]{BCL02} (the other versions yield comparable
results).  The setup is this: we oversample the signal in each
dimension by a factor of $r$, where $r=2,3,4,5$.  For each
oversampling rate, we run ten experiments using a different test
signal each time. The table shows the average residual errors over ten
runs as well as the average relative MSE. The ill-posedness of the
problem is evident from the disconnect between small residual error
and large reconstruction error; that is to say, we fit the data very
well and yet observe a large reconstruction error.  {\em Thus, in
  stark contrast to what is widely believed, our simulations indicate
  that oversampling by itself is not a viable strategy for phase
  retrieval even for non-negative, real-valued images.} 
\begin{table}[h]
\centering
  \begin{tabular}{|c|c|c|c|c|}
    \hline
    Algorithm $\vert$ Oversampling   & 2      & 3      & 4      & 5      \\
    \hline 
    \hline
    $\|\cA(X) - y\|_2/\|y\|_2$ (Fienup) & 0.0650 & 0.0607 & 0.0541 & 0.0713 \\
    \hline
    Relative MSE (Fienup) & 0.6931 & 0.6882 & 0.6736 & 0.6878 \\
    \hline 
    \hline
    $\|\cA(X) - y\|_2/\|y\|_2$ (PhaseLift) & 0.0051 & 0.0055 & 0.0056 & 0.0052 \\
    \hline
    Relative MSE (PhaseLift)   & 0.4932 & 0.4893 & 0.4960 & 0.4981 \\
    \hline
  \end{tabular}
  \caption{
    MSE obtained by Fienup's algorithm and by PhaseLift with reweighting
    from oversampled DFT measurements 
    taken on 2D real-valued and positive test images. 
    Fienup's algorithm does not always find a signal consistent 
    with the data as well as the support constraint. (After the projection step in the spatial domain, the current guess does not always match the measurement in Fourier space. After `projection' in Fourier space, the signal is not the Fourier transform of a signal obeying the spatial constraints.)
    Our approach always finds signals matching  
    measured data very well, and yet the reconstructions achieve a
    large reconstruction error. This indicates severe ill-posedness 
    since we have several distinct solutions providing an excellent 
    fit to the measured data.}
\label{tab:2C}
\end{table}

\section{Discussion}
\label{sec:discussion}

This paper introduces a novel framework for phase retrieval, combining
multiple illuminations with tools from convex optimization, which
been shown to work very well in practice and bears great
potential. Having said this, our work also calls for improved theory,
improved algorithms and a physical implementation of these
ideas. Regarding this last point, it would be interesting to design
physical experiments to test our methodology on real data, and we hope
to report on this in a future publication. For now, we would like to
bring up important open problems.

At the theoretical level, we need to understand for which families of
physically implementable structured illuminations does the trace-norm
heuristic succeed? How many diffraction patterns are provably
sufficient for our convex programming approach to work? Also, we have
shown that our approach is robust to noise in the sense that the
performance degrades very gracefully as the SNR decreases. Can this be
made rigorous? Can we prove that our proposed framework is indeed
robust to noise? Here, it is very likely that the tools and ideas
developed in the theories of compressed sensing and of matrix
completion will play a key role in addressing these fundamental
issues.

At the algorithmic level, we need to address the fact that the lifting
creates optimization problems of potentially enormous sizes. A
tantalizing prospect is whether or not it is possible to use
knowledge that the solution has low-rank, e.~g.~rank one, to design
algorithms which do not need to assemble or store very large matrices.
If so, how can this be done? Here, randomized algorithms holding up a
sketch of the full matrix may prove very helpful.

Finally, we would like to close by returning to another finding of
this paper. Namely, oversampling the Fourier transform---this is the
same as assuming finite support of the specimen---appears extremely
problematic in practice, even for real-valued nonnegative signals. To
be sure, we have demonstrated that there typically exist very distinct
2D signals whose modulus of the Fourier transform nearly coincide,
whatever the degree of oversampling. In light of this extreme ill
posedness, we have trouble understanding why this technique is used so
%heavily when it is shown to produce results of dubious scientific merit.  
heavily when it does not produce useful results in the absence of
very specific a priori information about the image.
Moreover, our concern is compounded by the additional fact
that algorithms in common use tend to return solutions that depend on
an initial guess so that different runs return widely different
solutions.  In the spirit of reproducible research, this calls for
documented results making publicly available both data sets and
software so that researchers can reproduce published results or
results yet to be published.  Of course, one might also be willing to
rely on image priors far stronger than finite spatial extent,
real-valuedness and positivity; they would, however, need to be
specified.

\begin{small}
\subsection*{Acknowledgements}

E.~C.\ is partially supported by NSF via grant CCF-0963835 and the
2006 Waterman Award. Y.~E.\ is partially supported by the Israel
Science Foundation under Grant no.\ 170/10. T.~S.\ acknowledges partial
support from the NSF via grants DMS 0811169 and DMS-1042939, and from
DARPA via grant N66001-11-1-4090. V.~V.\ is supported by the
Department of Defense (DoD) through the National Defense Science \&
Engineering Graduate Fellowship (NDSEG) Program. We are indebted to
Stefano Marchesini for inspiring and helpful discussions on the phase
problem in X-ray crystallography as well as for providing us with the
gold balls data set used in Section \ref{sec:numerics}. We want to
thank Philippe Jaming for bringing Wright's conjecture in
\cite{Vogt78} to our attention.

\bibliographystyle{plain}
\bibliography{phaserefs}
\end{small}

% \appendix
% \input{appendix}

\end{document}